Title: When is chemical disequilibrium in Earth-like planetary atmospheres a biosignature versus an anti-biosignature? Disequilibria from dead to living worlds.

Running Head: Chemical Disequilibrium as a Biosignature

Authors: Nicholas F. Wogan[1,2], David C. Catling[1,2]

Affiliations:
(1) Dept. Earth and Space Sciences, Box 351310, University of Washington, Seattle, WA
(2) Virtual Planetary Laboratory, University of Washington, Seattle, WA, 98195, USA.

Contact:
Nicholas Wogan
Department of Earth and Space Sciences
Box 351310, University of Washington
Seattle, WA 98195-1310
wogan@uw.edu

David Catling
Department of Earth and Space Sciences
Box 351310, University of Washington
Seattle, WA 98195-1310

When is chemical disequilibrium in Earth-like planetary atmospheres a biosignature versus an anti-biosignature? Disequilibria from dead to living worlds.

Nicholas F. Wogan[1, 2*], David C. Catling[1, 2]
[1]Department of Earth and Space Sciences/Astrobiology Program, University of Washington, Seattle, WA 98195, USA. [2]Virtual Planetary Laboratory, University of Washington, Seattle, WA, 98195, USA.
*Corresponding author. Email: wogan@uw.edu

**Abstract [249 words of a 250 limit]**

Chemical disequilibrium in exoplanetary atmospheres (detectable with remote spectroscopy) can indicate life. The modern Earth's atmosphere-ocean system has a much larger chemical disequilibrium than other solar system planets with atmospheres because of oxygenic photosynthesis. However, no analysis exists comparing disequilibrium on lifeless, prebiotic planets to disequilibrium on worlds with primitive chemotrophic biospheres that live off chemicals and not light. Here, we use a photochemical-microbial ecosystem model to calculate the atmosphere-ocean disequilibria of Earth with no life and with a chemotrophic biosphere. We show that the prebiotic Earth likely had a relatively large atmosphere-ocean disequilibrium due to the coexistence of water and volcanic $H_2$, $CO_2$, and CO. Subsequent chemotrophic life probably destroyed nearly all of the prebiotic disequilibrium through its metabolism, leaving a likely smaller disequilibrium between $N_2$, $CO_2$, $CH_4$, and liquid water. So, disequilibrium fell with the rise of chemotrophic life then later rose with atmospheric oxygenation due to oxygenic photosynthesis. We conclude that big prebiotic disequilibrium between $H_2$ and $CO_2$ or CO and water is an anti-biosignature because these easily metabolized species can be eaten due to redox reactions with low activation energy barriers. However, large chemical disequilibrium can also be a biosignature when the disequilibrium arises from a chemical mixture with biologically insurmountable activation energy barriers, and clearly identifiable biogenic gases. Earth's modern disequilibrium between $O_2$, $N_2$, and liquid water along with minor $CH_4$ is such a case. Thus, the interpretation of disequilibrium requires context. With context, disequilibrium can be used to infer dead or living worlds.

## 1. Introduction

It will soon be possible to look for biosignature gases in exoplanet atmospheres with telescopes. Within several years, the James-Webb Space Telescope (JWST) will measure the composition of exoplanet atmospheres with transit spectroscopy (Barstow & Irwin 2016) and, within decades, telescopes capable of reflectance spectroscopy will examine Earth-sized exoplanets around Sun-like stars (Fischer et al. 2019; Gaudi et al. 2019). Ground-based telescopes, such as the Extremely Large Telescope, will also play a role in the spectroscopic search for life by the mid 2020s (López-Morales et al. 2019; Snellen et al. 2013).

Much biosignature research suggests that telescopes look for $O_2$ produced by oxygenic photosynthesis (Meadows 2017; Meadows et al. 2018; Owen 1980). Molecular oxygen can be a relatively easy biogenic gas to detect on an exoplanet (Meadows 2017), and it is generated in large quantities by relatively few abiotic processes (Meadows, et al. 2018).

However, Earth's $O_2$ biosignature has been strongly detectable for only the past ~1/8th of Earth's inhabited history. Fossil stromatolites show that the origin of life was before ~3.5 Ga (Walter et al. 1980), while geochemical data suggest that oxygenic photosynthesis could have arisen by ~3 Ga (Planavsky et al. 2014a). Despite the possible early rise of oxygenic photosynthesis, there was negligible atmospheric $O_2$ in the Archean eon (4.0 to 2.5 Ga) (Farquhar et al. 2000). Earth had $O_2$ in the Proterozoic Eon (2.5 to 0.541 Ga), but some atmospheric proxies (Planavsky et al. 2014b) indicate that $O_2$ may not have been plentiful enough to detect over interstellar distances with upcoming and future space-based telescopes (Krissansen-Totton et al. 2018b; Reinhard et al. 2017). Also, oxygenic photosynthesis is a complex metabolism that only evolved once on Earth (Fischer et al. 2016), and it is unknown whether its origin on an exoplanet is likely.

An alternative to looking for any single biogenic gas (e.g., $O_2$, $CH_4$, or $N_2O$), is to look for chemical disequilibrium, i.e., the long-term coexistence of two or more chemically incompatible species (Lovelock 1965; Lovelock 1975). On the modern Earth, different metabolisms produce different waste gases, which have a thermodynamic drive to react over long periods of time. Thus, incompatible waste gases, or disequilibria, are maintained in Earth's environment by biogenic fluxes. The persistence of $CH_4$ and $O_2$ (which react through a series of intermediates) in Earth's modern atmosphere is an example and indicates continuous replenishment of these gases by biology.

Lovelock (1965) first proposed searching for life on other planets by looking for disequilibrium gases in planetary atmospheres, and subsequently Lovelock (1975) attempted to quantify the disequilibrium of Solar System planets. Unfortunately, knowledge of atmospheric composition of the Solar System planets, and computational methods for thermodynamic calculations were insufficient at the time for accurate calculations.

Using modern computational techniques and thermodynamics, Krissansen-Totton et al. (2016) calculated the atmosphere or atmosphere-ocean disequilibrium of several Solar System planets, Titan, and Earth. They found that Earth's atmosphere-ocean system has more than an order of magnitude disequilibrium (in joules per mole of atmosphere) than any other planet due to biogenic fluxes. They propose high atmosphere-ocean chemical disequilibrium as a biosignature for exoplanets similar to the modern Earth, with photosynthetic biospheres. Subsequently, Krissansen-Totton et al. (2018c) used atmospheric proxy and model-based estimates of Earth's Archean and Proterozoic atmosphere and ocean to calculate chemical disequilibrium over Earth history. They showed that disequilibrium rose to its present value because of atmospheric oxygen released from oxygenic photosynthesis, and $N_2$ put into the atmosphere from bacterial denitrification (conversion of $NO_x$ to $N_2$) which uses organic carbon from photosynthesis (for further explanation see Section 4.1 in Krissansen-Totton, et al. (2016)).

Despite this prior work, interpretation of disequilibrium as a sign of life is unclear. A planet without life might have a large disequilibrium of untapped free energy because life is not consuming it, so disequilibrium in that case is the very opposite of a sign of life: an anti-biosignature. If chemotrophic life evolves, its metabolism uses environmental free energy and tends to push environments toward thermodynamic equilibrium. Thus, we expect no big disequilibrium on a purely chemotrophic world. Finally, the modern state of high disequilibrium *is* a biosignature, but depends on the presence of a large, oxygenic photosynthetic biosphere.

To elucidate these subtleties quantitatively, we use a photochemical model to calculate the plausible atmosphere-ocean disequilibrium of the prebiotic Earth and then couple the model to a simple microbial biosphere to investigate the Earth immediately after the origin of life. We demonstrate that atmosphere-ocean disequilibrium drops when chemotrophic life appears because such life destroys volcanically generated atmospheric free energy and can easily catalyze the reactions. However, the mixture of gases from phototrophs is not all consumed by chemotrophs because of insurmountable activation energy barriers, so this disequilibrium persists. Our results build upon previous studies (Krissansen-Totton, et al. 2016; Krissansen-Totton, et al. 2018c) to provide conservative estimates of chemical disequilibrium through Earth history by including the Hadean Earth. With our results, we distinguish the general cases when disequilibrium indicates life versus when disequilibrium is an anti-biosignature.

## 2. Methods

We model the change in atmosphere-ocean chemical disequilibrium between the prebiotic Earth, and Earth influenced by a chemotrophic ecosystem in two steps. First, we simulate atmospheric composition using a photochemical model coupled to a microbial biosphere (in the biotic case), and second, we calculate the atmosphere-ocean disequilibrium of this simulated atmosphere with multiphase Gibbs energy minimization. The following sections briefly describe both of these steps, and the Appendices A and B contain more detailed methods. The Python, Fortran and MATLAB source code is available on Github at https://github.com/Nicholaswogan/Wogan_and_Catling_2020.

### 2.1. Modeling the Hadean Atmosphere

For both the prebiotic and biotic atmospheric compositions, we use the 1-D photochemical-climate code contained within the open source software package *Atmos*. *Atmos* is derived from a model originally developed by the Kasting group (Pavlov et al. 2001), and versions of this code have been used to simulate the Archean and Proterozoic Earth atmosphere (Zahnle et al. 2006), Mars (Sholes et al. 2019; Smith et al. 2014; Zahnle et al. 2008), and exoplanet atmospheres (Arney et al. 2016; Schwieterman et al. 2019). We use *Atmos* to model the prebiotic atmosphere and the atmosphere influenced by a chemotrophic ecosystem by setting lower boundary conditions appropriate to each scenario. Every model run achieves redox balance (i.e., conservation of chemical oxidants and reductants in the atmosphere) to better than approximately 0.01% (for an explanation of redox balance see Chapter 8 in Catling and Kasting (2017)).

#### *2.1.1. Hadean Volcanic Outgassing*

Modeling the atmosphere requires estimates of volcanic outgassing fluxes on the Hadean Earth. These fluxes depend on the redox state of the mantle, which is quantified by the mantle's oxygen fugacity ( $f_{O_2}$ ). A more reduced mantle (lower $O_2$ fugacity) expels more reduced gases (e.g., $H_2$) relative to oxidized gases (e.g., $H_2O$). Recent oxygen fugacity proxies indicate that Earth's mantle was more reduced several billion years ago and slowly oxidized (Aulbach & Stagno 2016; Nicklas et al. 2019). We linearly extrapolate $O_2$ fugacity data obtained by Aulbach and Stagno (2016) backward in time to estimate an $O_2$ fugacity of $\log(f_{O_2}) = \mathrm{FMQ} - 1.48$ at ~4 Ga (Appendix A) to represent mantle redox state around the time of the origin of life. Here, FMQ is the fayalite–magnetite-quartz buffer which is a synthetic reference $f_{O_2}$ value at fixed temperature-pressure conditions. Sensitivity of calculated disequilibrium to $f_{O_2}$ is relatively small. Changing the oxygen fugacity by 1 log unit changes the calculated atmosphere-ocean chemical disequilibrium by a factor of ~2 (Appendix B.3), which is small compared to other uncertainties in chemical disequilibrium for an assumed prebiotic Earth at 4 Ga.

Volcanic outgassing in prebiotic times also depends on the total flux of all volcanic gases. This total depends on the tectonic regime of the ancient Earth, which is debated (Rosas & Korenaga 2018). If Earth lacked plate tectonics and was in a "stagnant lid" regime, then the average heat flux could have been comparable to the modern flux despite a much warmer interior (Korenaga 2009). On the other hand, if plate tectonics was active in the Hadean, the heat flux on the 4 Ga Earth could have been 5 times higher than today (Zahnle et al. 2001).

Volcanic outgassing is proportional to the heat flux to a power between 1 and 2. To be conservative, we take volcanic outgassing proportional to the square of heat flux (Sleep & Zahnle 2001), so lower and upper bounds on heat flux suggest volcanic outgassing rates between 1 and 25 times the modern outgassing rate. We adopt this range here to estimate total volcanic outgassing fluxes ( $F_x$ ) of hydrogen, carbon and sulfur at ~4 Ga with the formulas

$$F_{\text{hydrogen}} = CF_{\text{hydrogen}}^{\text{mod}} \tag{1}$$

$$F_{\text{carbon}} = CF_{\text{carbon}}^{\text{mod}} \tag{2}$$

$$F_{\text{sulfur}} = CF_{\text{sulfur}}^{\text{mod}} \tag{3}$$

Here, $F_x^{\text{mod}}$ is the modern outgassing flux of species $x$, and $C$ is an outgassing multiplier that we vary between 1 and 25. Fluxes are calculated in units of molecules $cm^{-2}$ $s^{-1}$.

With estimates of $O_2$ fugacity and total outgassing fluxes, we use equilibrium chemistry of the mantle to calculate plausible outgassing fluxes of individual gases, $H_2$, $H_2O$, $CH_4$, $CO_2$, CO, $H_2S$, and $SO_2$ for $C$ between 1 and 25. Details of these calculations are in Appendix A.

### *2.1.2. Modeling a Prebiotic Atmosphere*

We model the Earth's prebiotic atmosphere for each volcanic outgassing scenario between 1 and 25 times modern outgassing. We use calculated outgassing fluxes of $H_2$, CO, $SO_2$, and $H_2S$ as lower boundary conditions to the *Atmos* photochemical model. Additionally, we set a CO deposition velocity to $1.0\times10^{-8}$ cm $s^{-2}$ to reflect the abiotic uptake of CO by the ocean (Kharecha et al. 2005). We assume that the abiotic surface flux of $CH_4$ is negligible. This assumption is supported by a recent work on the abiotic formation $CH_4$ on the modern Earth (Fiebig et al. 2019) but is disputed by other studies (Etiope & Sherwood Lollar 2013). All other boundary conditions are specified in Appendix B.1. Given volcanic outgassing fluxes and other boundary conditions, *Atmos* calculates the mixing ratios of all species when the atmosphere is at photochemical equilibrium.

### *2.1.3. Modeling an Atmosphere Influenced by a Chemotrophic Ecosystem*

For each volcanic outgassing scenario, we also model atmospheric composition influenced by a marine ecosystem of chemotrophic microbes. Our oceanic ecosystem is composed of four chemotrophic microorganisms with the following metabolisms:

$$CO_2 + 4H_2 \rightarrow CH_4 + 2H_2O \tag{4}$$

$$2CH_2O \rightarrow CH_3COOH \tag{5}$$

$$CH_3COOH \rightarrow CH_4 + CO_2 \tag{6}$$

$$4CO + 2H_2O \rightarrow 2CO_2 + CH_3COOH \tag{7}$$

These equations represent the metabolisms of chemosynthetic methanogens (Equation (4)), acetogenic bacteria (Equation (5)), acetotrophic methanogens (Equation (6)), and CO-consuming acetogens (Equation (7)). We have chosen this ecosystem to represent Earth's biosphere after the origin of life and before the origin of photosynthesis. The actual make-up Earth's biosphere at this time is unknown, but all organisms in our chosen ecosystem are phylogenetically ancient and should have preceded photosynthesis (Adam et al. 2018; Schönheit et al. 2016; Wolfe & Fournier 2018), so they are a reasonable representation.

We model the impact of these various organism on atmospheric composition by setting lower boundary conditions in the photochemical model that reflect their metabolisms. This technique was used by Kharecha, et al. (2005), and our ecosystem model is nearly identical to their "case 2" atmosphere-ecosystem model. The only difference is that the *Atmos* photochemical code is an updated version of the one used by Kharecha, et al. (2005). Below, we briefly describe how the model works, although a more in-depth account can be found in Kharecha, et al. (2005) p. 58-61. Appendix B.1 contains all the boundary conditions that are not listed in the main text.

Ground-level $H_2$ was likely much more plentiful than $CH_4$ on the prebiotic Earth because $H_2$ was produced by mantle-sourced volcanoes, and $CH_4$ was not because it is not thermodynamically favored compared to $CO_2$. When chemotrophic methanogens originated, they would have converted some of the prebiotic $H_2$ to $CH_4$ through their metabolism, although the total amount of hydrogen stored in these molecules would not have changed significantly. In other words, the weighted sum of the ground-level $H_2$ and $CH_4$ mixing ratios on the prebiotic Earth (denoted $n_{\mathrm{H_2}}^{\mathrm{pre}}$ and $n_{\mathrm{CH_4}}^{\mathrm{pre}}$, respectively) would have been approximately equal to the weighted sum of the ground-level $H_2$ and $CH_4$ mixing ratios on the Earth influenced by methanogens (denoted $n_{\mathrm{H_2}}^{\mathrm{eco}}$ and $n_{\mathrm{CH_4}}^{\mathrm{eco}}$, respectively):

$$n_{\mathrm{H_2}}^{\mathrm{eco}} + 2n_{\mathrm{CH_4}}^{\mathrm{eco}} \approx n_{\mathrm{H_2}}^{\mathrm{pre}} + 2n_{\mathrm{CH_4}}^{\mathrm{pre}} \tag{8}$$

Equation (8) is only approximately valid because burial of organic carbon, which contains hydrogen, would cause $n_{\mathrm{H_2}}^{\mathrm{eco}} + 2n_{\mathrm{CH_4}}^{\mathrm{eco}}$ to be less than $n_{\mathrm{H_2}}^{\mathrm{pre}} + 2n_{\mathrm{CH_4}}^{\mathrm{pre}}$ by no more than ~1%. The precise difference depends on how efficiently organic carbon was buried in the past. Since this difference is small, we ignore organic carbon is burial, and assume that acetogenic bacteria and acetotrophic methanogens living in the ocean convert all organic carbon to methane and carbon dioxide. Our assumptions implicitly include heterotrophs in the model.

How much of the prebiotic atmospheric $H_2$ was converted to $CH_4$ by methanogens? Methanogens lived in the ocean, so their consumption or generation of atmospheric $H_2$ and $CH_4$ was modulated by gas transfer across the atmosphere-ocean interface. We model gas exchange using a stagnant boundary layer model (Kharecha, et al. 2005; Liss & Slater 1974). Within the ocean, life was probably energy limited, and not nutrient limited (i.e., life was not limited by phosphorus or biologically available nitrogen) on Earth before the advent of oxygenic photosynthesis (Canfield et al. 2006; Kharecha, et al. 2005; Ward et al. 2019). Therefore, we assume that methanogens consumed $H_2$ and expelled $CH_4$ in the ocean until they obtain 30 kJ mol$^{-1}$ from Equation (4), which is the approximate Gibbs energy required to create 1 mol of ATP.

In practice, we simulate methanogens for each outgassing rate with the following steps. First, we arbitrarily set the ground-level $H_2$ and $CH_4$ mixing ratios in the photochemical model such that they satisfy Equation (8). Second, we run the photochemical model to retrieve the surface flux of $H_2$ and $CH_4$. Third, we check whether the calculated $H_2$ and $CH_4$ fluxes reflect energy-limited methanogens in an ocean which exchanges gases with the atmosphere via a stagnant boundary layer. Fourth, if the fluxes do not satisfy this requirement, then we select new $H_2$ and $CH_4$ mixing ratios which are closer to satisfying step 3 (which still satisfy Equation (8)). We iterate steps 2 through 4 until step 3 is satisfied.

To simulate CO-consuming acetogens, we set the CO deposition velocity to its maximum value of $1.2\times10^{-4}$ cm s$^{-1}$. This maximum deposition velocity assumes that acetogens consume CO as soon as it enters the ocean. This assumption is reasonable because an energy limited chemotrophic biosphere which contains CO consumers should draw CO concentrations to negligible amounts in the ocean (Kharecha, et al. 2005; Schwieterman, et al. 2019). The photochemical code calculates the mixing ratio of CO corresponding to the maximum deposition velocity.

## 2.2. Quantification of Chemical Disequilibrium

For each prebiotic and biotic atmosphere, we calculate the atmosphere-ocean chemical disequilibrium with Gibbs energy minimization, using code described previously (Krissansen-Totton, et al. 2018c). Given the chemical composition of an atmosphere-ocean system, the code reacts all molecules and atoms to thermodynamic equilibrium. The chemical disequilibrium is then defined by the Gibbs free energy difference between the initial and equilibrium state:

$$\Phi \equiv G_{(T,P)}(\mathbf{n}_{\text{initial}}) - G_{(T,P)}(\mathbf{n}_{\text{final}}) \qquad (9)$$

Here, $\Phi$ is the available Gibbs energy (J/mol atmosphere). The vector containing the abundance of all atmospheric and ocean species is $\mathbf{n}_{\text{initial}}$, while $\mathbf{n}_{\text{final}}$ contains abundances of the final equilibrium state. The quantification of chemical disequilibrium, $\Phi$, is the maximum chemical energy that can be extracted from the atmosphere-ocean system that can be used to do work.

We determined the initial state of the atmosphere using the surface mixing ratios from the photochemical model (as described in the previous two sections), while the assumed initial state of the ocean is given in Table 1. Unless stated otherwise in Table 1, dissolved gas abundances were determined with Henry's law constants derived from NASA's thermodynamic database (Burcat & Ruscic 2005) and SUPCRT database (Johnson et al. 1992). Additionally, we assumed atmospheric temperature and pressure to be 25°C and 1 bar respectively. Chemical disequilibrium is fairly insensitive to ocean composition, atmospheric pressure and temperature (Krissansen-Totton, et al. 2018c); consequently, order of magnitude errors in these assumptions will result in a fairly small error (well within a factor of ~2) in the available Gibbs energy.

**Table 1**

Assumed initial atmosphere-ocean composition for the prebiotic and biotic early Earth.

| Ocean Species | Molality (mol/kg) | Reference/explanation |
|---|---|---|
| $Na^+$ | 0.586 | Charge balance |
| $Cl^-$ | 0.545 | Modern value |
| $SO_4^{2-}$ | 0 | (Crowe et al. 2014) |
| $NH_3$ | 6.40E-09 | Henry's law from atmospheric $NH_3$ |
| $NH_4^+$ | 2.9E-06 | Equilibrium with $NH_3$ and pH |
| $H_2S$ | 0 | (Krissansen-Totton, et al. 2018c) |
| pH | 6.6 (dimensionless) | (Krissansen-Totton et al. 2018a) |
| $HCO_3^-$ | 0.02674 | Equilibrium with $CO_2$ and pH |
| $CO_3^{2-}$ | 8.03E-05 | Equilibrium with $HCO_3^-$ and pH |

| Atmospheric Species | Mixing Ratio | Reference/explanation |
|---|---|---|
| $NH_3$ | 1.00E-10 | Wolf and Toon (2010). Negligible, so not in photochemical model |
| $H_2O$ | 0.025 | Global average value |
| $CO_2$ | 0.2 | Nominal value (Kadoya et al. 2020; Krissansen-Totton, et al. 2018a) |

## 3. Results

### 3.1. Chemical disequilibrium on the prebiotic and chemotrophic Earth

The modeled mixing ratios of $H_2$, $CH_4$ and CO for both prebiotic and chemotrophic simulations are shown in Figure 1 as a function of the volcanic outgassing multiplier (from Equations (1) - (3)). All mixing ratios increase with increased volcanic outgassing, and CO in the prebiotic atmosphere increases rapidly. This behavior has been observed in other photochemical modeling studies and has been termed "CO runaway" (Kasting et al. 1983; Zahnle 1986). The CO consumers in the chemotrophic model prevent "CO runaway". Additionally, >95% of the $H_2$ present in the prebiotic model is converted to $CH_4$ by methanogens once we implement the chemotrophic model.

Figure 2 shows the modeled atmosphere-ocean thermodynamic disequilibrium for the prebiotic and chemotrophic atmosphere as a function of the volcanic outgassing multiplier. For all outgassing scenarios, the chemotrophic atmosphere-ocean disequilibrium is lower than the prebiotic atmosphere-ocean disequilibrium because the biosphere exploits free energy for metabolism. Additionally, the atmosphere-only disequilibrium is always lower in the chemotrophic ecosystem than in the prebiotic ecosystem for the same reason.

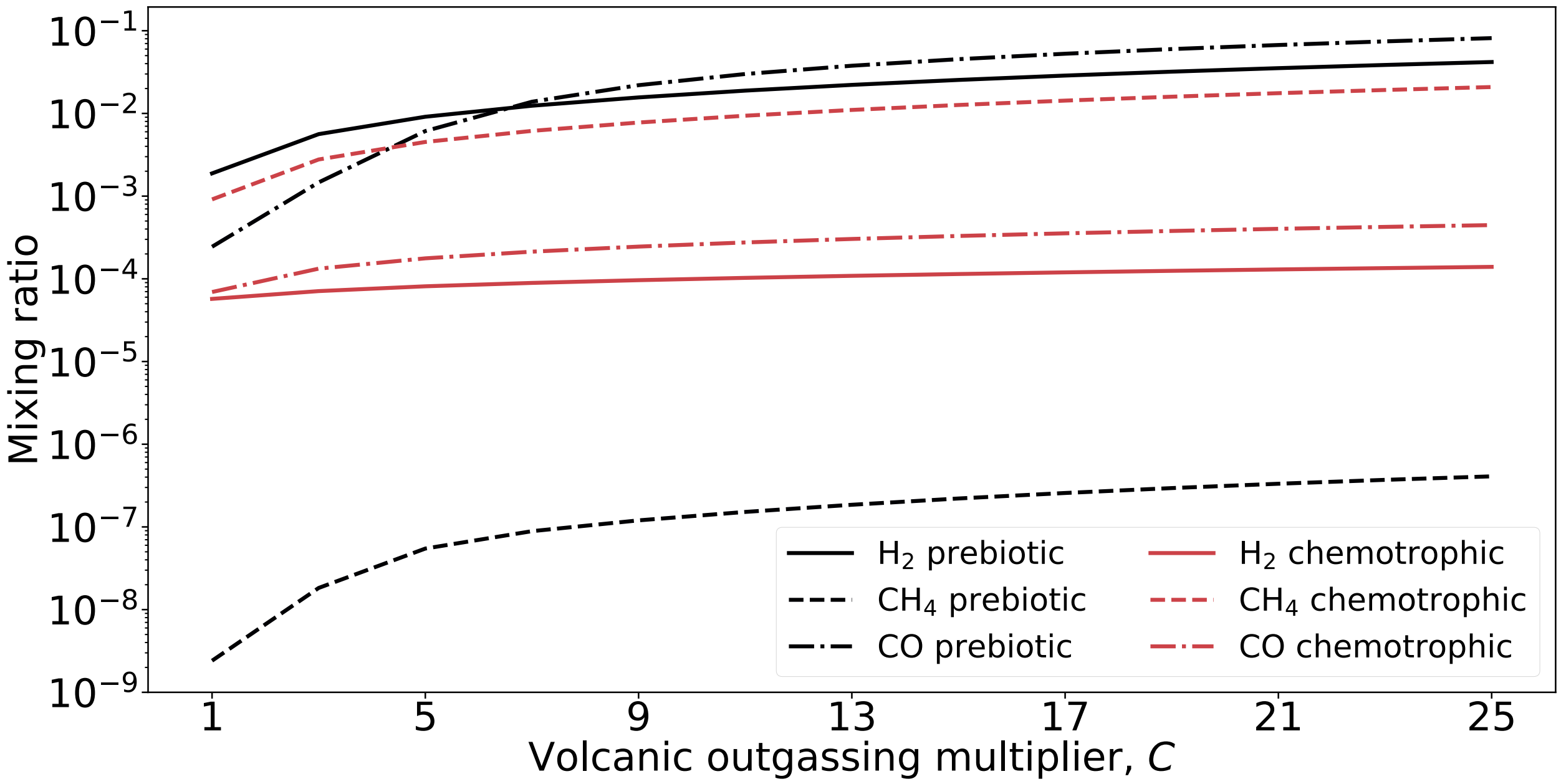

Figure 1: The mixing ratio of $H_2$, $CH_4$ and CO in the modeled prebiotic and chemotrophic early Earth atmospheres as a function of volcanic outgassing, relative to modern. Black lines represent mixing ratios for the prebiotic case. Red lines represent mixing ratios for the chemotrophic case where we have assumed an energy-limited ocean ecosystem. For both simulations, we also assume the mixing ratios of $N_2$ and $CO_2$ are 0.75 and 0.2 respectively. The presence of a chemotrophic biosphere drastically lowers $H_2$ abundances and increases $CH_4$ abundances due to methanogenesis, and lowers CO abundances because of acetogenesis.

The following sections explain which species contribute most to the available Gibbs energy in both the prebiotic and chemotrophic model.

## 3.2. The prebiotic disequilibrium and the species that contribute to it

The available Gibbs energy of the prebiotic atmosphere-ocean system for modern volcanic outgassing rates ( $C = 1$) is 62 J/mol of atmosphere (compared to 2326 J/mol for the modern biotic Earth (Krissansen-Totton, et al. 2016)). The largest source of disequilibrium is due to the coexistence of $CO_2$ and $H_2$ which accounts for ~40 J/mol (65%) of this total available Gibbs energy. These molecules should react and form $CH_4$ and water vapor in equilibrium:

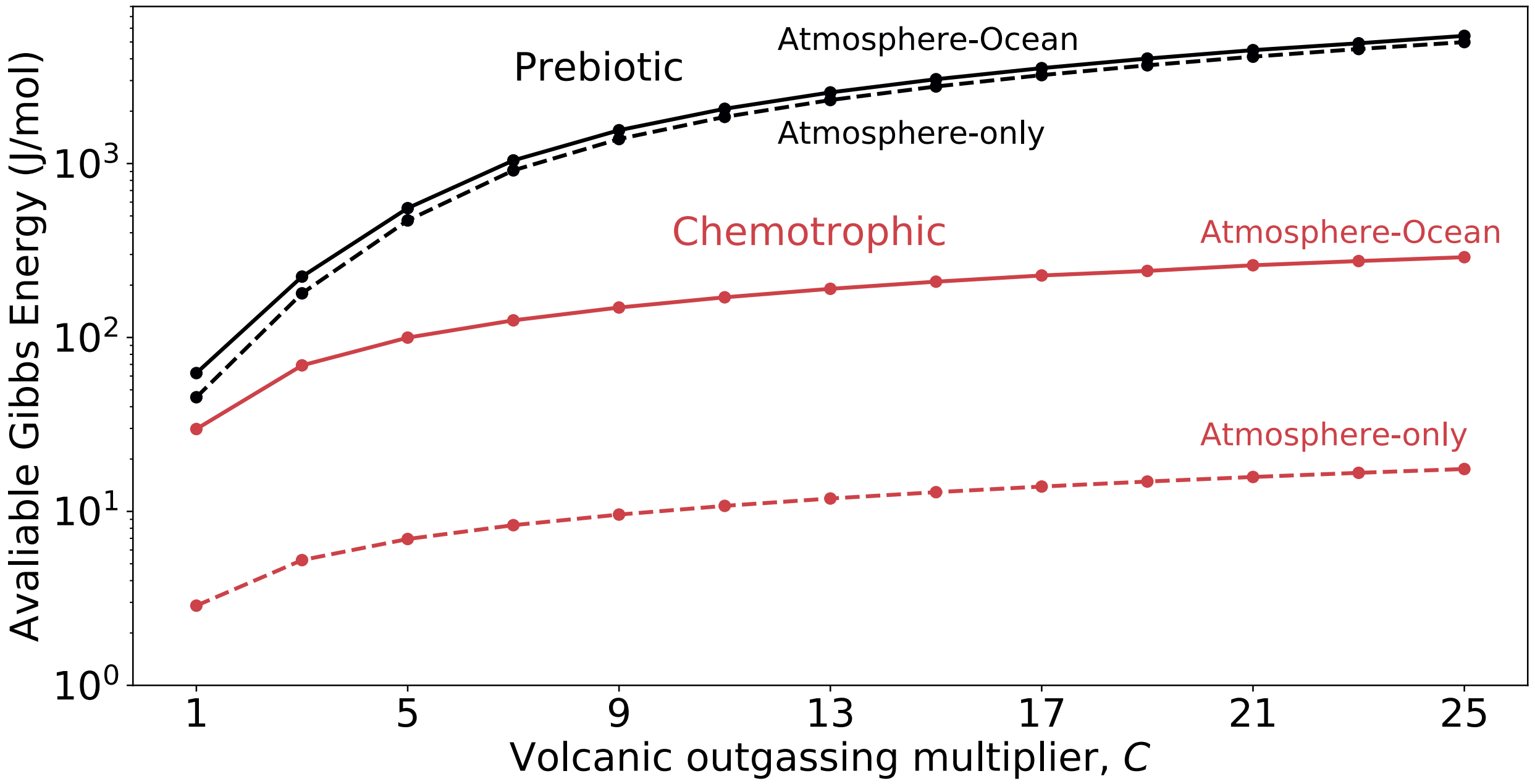

Figure 2: Chemical disequilibrium, as measured by available Gibbs energy, of the prebiotic (black lines) and chemotrophic (red lines) Earth as a function of a volcanic outgassing multiplier, relative to modern. The dotted lines are atmosphere-only Gibbs energy calculations, and the solid lines are atmosphere-ocean calculations. The presence of a chemotrophic ecosystem lowers both the atmosphere-ocean and atmosphere-only chemical disequilibrium by using the free energy for metabolism.

$$4H_2 + CO_2 \rightleftarrows CH_4 + 2H_2O \quad (10)$$

The coexistence of CO and water vapor contributes ~10 J/mol (16%), which is the second most important contributor to this available Gibbs energy. At equilibrium, $H_2$ and $CO_2$ will be replaced by $CH_4$ and $CO_2$ from the reaction

$$4CO + 2H_2O \rightleftarrows 3CO_2 + CH_4 \tag{11}$$

Both the $H_2$-CO and CO-$H_2O$ disequilibrium ultimately come from volcanic outgassing. Gases were once in equilibrium with magma but have been emitted into a colder environment of the atmosphere where they are in disequilibrium. For higher outgassing scenarios, the $H_2$-$CO_2$ and CO-$H_2O$ reactions remain the most import contributors to the available Gibbs energy. Since these reactions are in the gas phase, the atmosphere-only disequilibrium is nearly as large (~80%) as the atmosphere-ocean disequilibrium for all outgassing rates. For a possible Hadean outgassing rate of $C = 9$, $\mathbf{\Phi}$ is 1555 J/mol.

## 3.3. The chemotrophic disequilibrium and species that contribute to it

The atmosphere-ocean available Gibbs energy of the chemotrophic Earth for modern volcanic outgassing rates ( $C = 1$) is 30 J/mol. The coexistence of $CO_2$, $CH_4$, $N_2$, and liquid water contribute ~24 J/mol (80%) to this available Gibbs energy. These four species should react and deplete 99.9% of atmospheric methane in equilibrium

$$5CO_2 + 4N_2 + 3CH_4 + 14H_2O \rightleftarrows 8NH_4^+ + 8HCO_3^- \tag{12}$$

For volcanic outgassing 25 times modern fluxes ( $C = 25$), this reaction accounts for ~273 J/mol (94%) of the available Gibbs energy (290 J/mol), which shows that these species are the most important for all modeled chemotrophic systems. The atmosphere-only disequilibrium is always much smaller than the atmosphere-ocean disequilibrium because Equation (12) involves disequilibrium with the liquid water ocean.

The $H_2$-$CO_2$ and CO-$H_2O$ disequilibria, which dominate the prebiotic available Gibbs energy, contribute only ~0.8 J/mol and ~2.4 J/mol, respectively, for modern volcanic outgassing ( $C = 1$). The minor contribution of these disequilibria persists for all volcanic outgassing scenarios.

## 3.4. Disequilibrium though Earth history

Figure 3 shows our estimates of the evolution of Earth's atmosphere-ocean and atmosphere-only disequilibrium through its history. The prebiotic and chemotrophic disequilibrium ranges are from this study (i.e., Figure 2), and the estimates from the late Archean to the present are from Krissansen-Totton, et al. (2018c). Figure 3 has a broken axis between the chemotrophic ecosystem and the Archean because the advent of anoxygenic photosynthesis would have likely influenced how disequilibrium changed between these two eras. Our modeling does not capture this transition for reasons discussed below.

Like the chemotrophic Earth, the Archean disequilibrium was dominated by the coexistence of $CO_2$, $CH_4$, $N_2$, and liquid water (Krissansen-Totton, et al. 2018c). After the Great Oxidation Event, the magnitude of the available Gibbs energy rose, and was instead dominated by the coexistence $O_2$, $N_2$ and liquid water, which should react to form nitric acid at equilibrium:

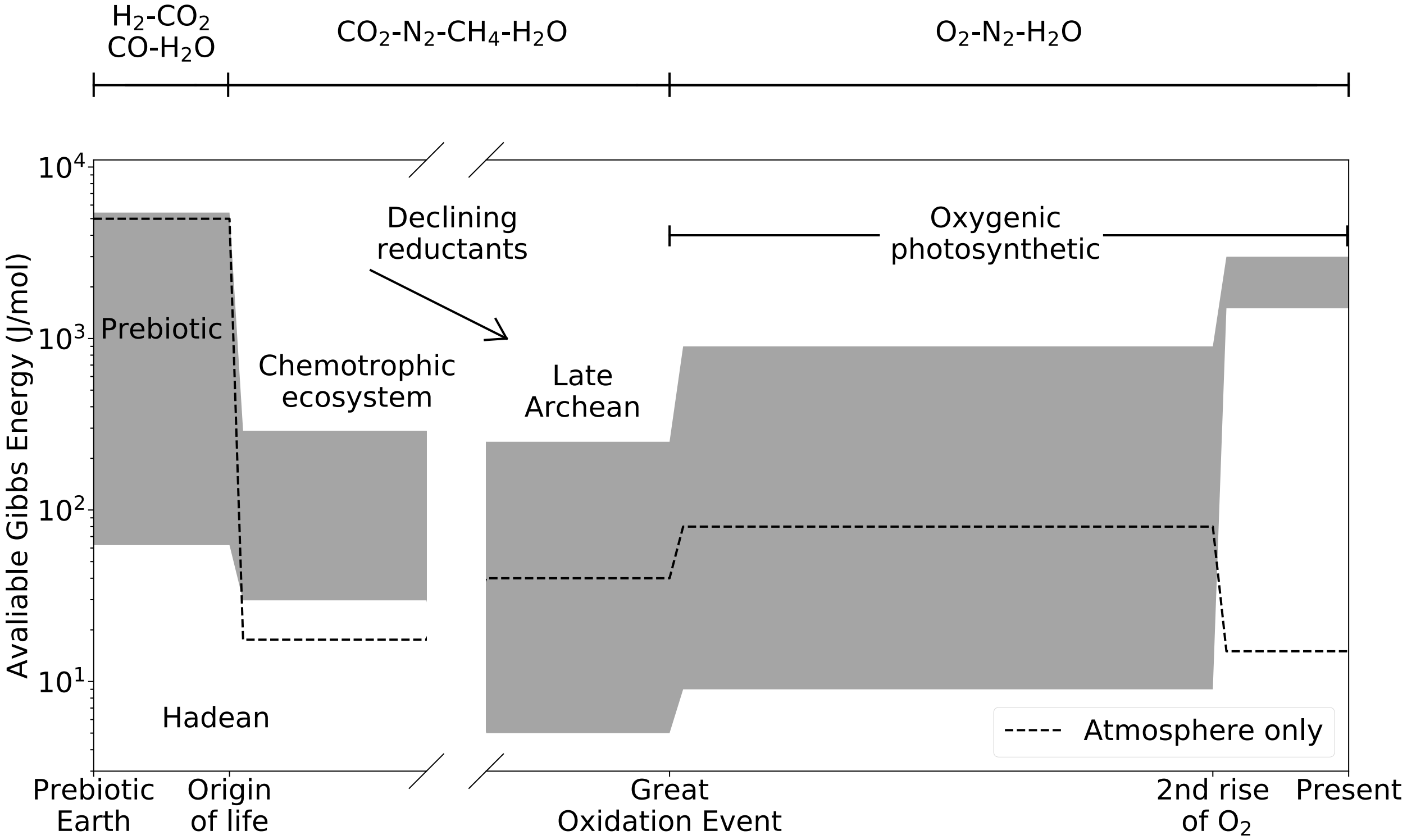

Figure 3: The chemical disequilibrium of Earth's atmosphere-ocean system through time. The shading indicates plausible ranges of atmosphere-ocean disequilibrium during intervals of Earth's history based on modeling (this study), and atmospheric proxies and models (Krissansen-Totton, et al. 2018c). The plot is broken between the "chemotrophic ecosystem" and "Archean" because the advent of anoxygenic photosynthesis would have likely influenced how disequilibrium changed between these two eras which is uncertain. The dotted line is the maximum atmosphere-only disequilibrium. Above the plot are the disequilibria (e.g., $H_2$-$CO_2$)

that contribute most to the atmosphere-ocean available Gibbs energy. Throughout Earth's history, disequilibrium fell with the rise of chemotrophic life, and rose after of the oxygenation of Earth's atmosphere from oxygenic photosynthesis.

$$5O_2 + 2N_2 + 2H_2O \rightleftarrows 4H^+ + 4NO_3^- \quad (13)$$

The magnitude of the $O_2$-$N_2$-$H_2O$ disequilibrium increased with the rise of $O_2$ until the present available Gibbs energy of 2326 J/mol (Krissansen-Totton, et al. 2016).

## 4. Discussion

### 4.1. Life's impact on disequilibria through Earth's history

Our results show that life has both generated and destroyed chemical disequilibrium in Earth's atmosphere-ocean system (Figure 3). Pioneering work by Lovelock (1975), which proposed using disequilibrium as a sign of life, argued that abiotic worlds would be close to thermodynamic equilibrium. However, this thinking ignored volcanically active planets. We showed that disequilibrium was likely high ($10^2$ to $10^3$ J/mol) in prebiotic times due to the volcanically produced $H_2$-$CO_2$ and CO-$H_2O$ disequilibria.

Subsequently, if the first life was chemotrophic and metabolized $H_2$, $CO_2$, and CO, then the atmosphere-ocean disequilibrium would have dropped to ~$10^2$ J/mol with the rise of microbial life. This is an example of chemotrophic life destroying the disequilibrium of its environment and promoting chemical equilibrium on a global scale.

The invention of anoxygenic photosynthesis, which we did not consider, may have added to the Atmosphere-ocean disequilibrium in the late Archean. Iron oxidizing photosynthesis is an example:

$$4Fe^{2+} + CO_2 + 11H_2O + h\nu \rightarrow 4Fe(OH)_3 + CH_2O + 8H^+ \quad (14)$$

The $CH_2O$ produced could have been processed by heterotrophs and methanogens yielding $CH_4$, which would have added to the Archean $CO_2$-$N_2$-$CH_4$-$H_2O$ disequilibrium without the need for additional volcanic outgassing (Krissansen-Totton, et al. 2018c). Additionally, the $CH_2O$ would also degrade into CO in the ocean, which would have added a small amount to the CO-$H_2O$ disequilibrium (Schwieterman, et al. 2019). Figure 3 does not explicitly capture these effects because the evolutionary history of anoxygenic photosynthesis is uncertain, but Archean disequilibrium estimates allow for such photosynthesis (Krissansen-Totton, et al. 2018c).

Even though the rise of anoxygenic photosynthesis would have added to the Late Archean disequilibrium, overall disequilibrium may have dropped because a lower flux of reductants would have been available to the biosphere. Before the rise of oxygenic photosynthesis, which uses ubiquitous water and sunlight, the biosphere is hypothesized to have been probably limited by the available reductants such as $H_2$, $Fe^{2+}$, and CO (Canfield, et al. 2006). For example, $H_2$-using anoxygenic phototrophs ( $CO_2 + 2H_2 + h\upsilon \rightarrow CH_2O + H_2O$ ) were likely limited by volcanically outgassed $H_2$. Volcanic outgassing of reductants probably declined from the Hadean to the late Archean as the Earth's interior cooled. Fewer available reductants would have lowered biological $CH_4$ production, resulting in smaller disequilibrium in the late Archean.

The increase of the available Gibbs energy and the rise of the $O_2$-$N_2$-$H_2O$ disequilibrium after the Great Oxidation Event was primarily caused by oxygenic photosynthesis. Atmospheric $O_2$ comes directly from oxygenic photosynthesis, and $N_2$ is generated, in part, from denitrifying bacteria that are ultimately powered by organic material from photosynthesis. Disequilibrium increased again to near modern levels with a rise of $O_2$ to near modern levels through the Neoproterozoic and Paleozoic (Krause et al. 2018).

### 4.2. Why disequilibrium persists in Earth's atmosphere-ocean system despite the presence of biology

Chemotrophs consumed a large fraction of Earth's prebiotic disequilibrium (Figure 2), but microbes left the $CO_2$-$N_2$-$CH_4$-$H_2O$ and $O_2$-$N_2$-$H_2O$ disequilibrium uneaten in the Archean and Proterozoic eons and in modern times. Thus, a pertinent question is: Why didn't microbes evolve metabolisms to consume the "free lunch" that has persisted in Earth's atmosphere?

We propose that this lack of consumption is due to the kinetic barriers of the $CO_2$-$N_2$-$CH_4$-$H_2O$ and $O_2$-$N_2$-$H_2O$ reactions, which we hypothesize are insurmountable by enzymes. To illustrate this idea, consider the disequilibrium of $O_2$-$N_2$-$H_2O$. These species would react slowly in the atmosphere in the absence of life via a number of steps:

$$
\begin{aligned}
2N_2 + 2O &\rightarrow 2NO + 2N \\
2N + 2O_2 &\rightarrow 2NO + 2O \\
4NO + 2O_2 &\rightarrow 4NO_2 \\
4NO_2 + O_2 + 2H_2O &\rightarrow 4HNO_3 \\
4HNO_3 &\rightarrow 4H^+ + 4NO3^-
\end{aligned}
\tag{15}
$$

The first two reactions, which make NO, are Zeldovich's reactions (Dixon-Lewis 1984) and require lightning to heat the air to ~20,000 K (Chameides et al. 1977). The third reaction occurs very quickly after the NO is generated (Murray 2016). The final two reactions are ultimately (partially) responsible for acid rain (Platt 1986). The rate limiting step to the net reaction is the first one, which has an activation energy of 316 kJ/mol (Dixon-Lewis 1984). We take this to be a lower bound on the uncatalyzed activation energy of reacting $O_2$, $N_2$ and $H_2O$. This must be a lower bound because the rate limiting step requires the presence of atomic oxygen, which could only have come from splitting $O_2$ with additional energy.

Life harnesses the free energy of disequilibria by lowering activation energy barriers with enzymes. Figure 4a is a classic textbook graph of free energy during an exothermic chemical reaction. Uncatalyzed reactions can only occur if a relatively large activation energy barrier is overcome. Therefore, many uncatalyzed reactions (between disequilibria) occur extremely slowly because ambient thermal energy is insufficient. Microbes tap into the free energy stored in disequilibria by using enzymes to lower activation energy barriers to levels where thermal energy allows reactions to proceed at appreciable rates.

Figure 4b compares the uncatalyzed activation energy of $O_2$-$N_2$-$H_2O$ to the uncatalyzed activation energy (blue bars) of reactions that enzymes lower to ~30 to 60 kJ/mol, which allow reactions to proceed at normal temperatures. The reaction between $O_2$, $N_2$, and $H_2O$, which is not performed by life, has an activation energy that is higher than all other uncatalyzed reactions. This suggests that Reaction (13) is not amenable to biological catalysis. The activation energy of which we hypothesize explains why no life has evolved that can exploit the $O_2$-$N_2$-$H_2O$ disequilibrium.

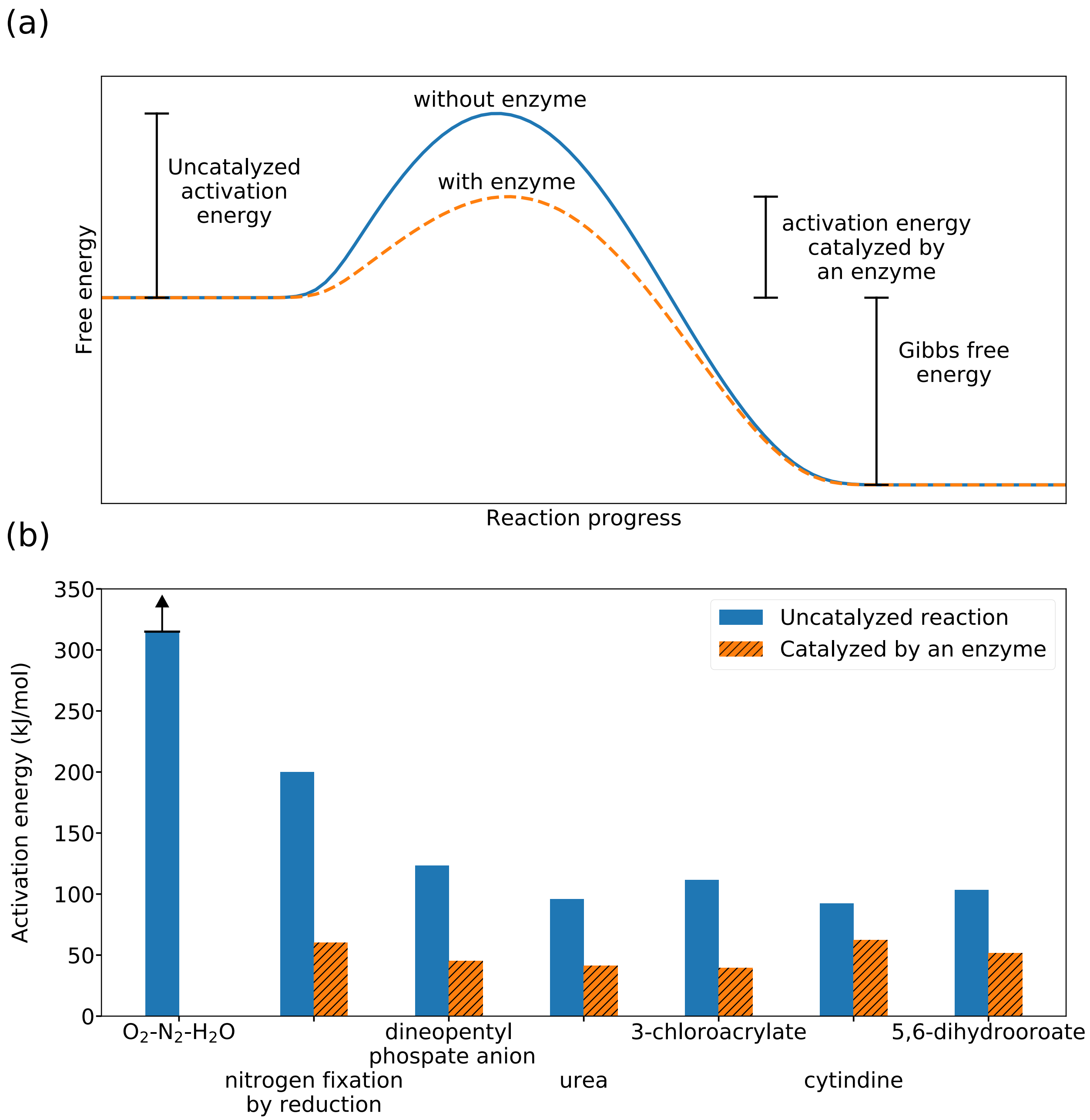

Figure 4: (a) Schematic of free energy during a chemical reaction. (b) The activation energy of several uncatalyzed reactions (blue), and reactions catalyzed by enzymes (orange). The lower bound for the uncatalyzed activation energy of $O_2$-$N_2$-$H_2O$ (a reaction that life doesn't perform) is from Dixon-Lewis (1984), and the activation energy of nitrogen fixation is from a number of references (Andersen & Shanmugam 1977; Hageman & Burris 1980) (see Appendix C for a summary of our literature search of nitrogen fixation kinetics). The rest of the activation energies are from Table 4 in Wolfenden (2006). The uncatalyzed activation energy of $O_2$-$N_2$-$H_2O$ is notably larger than the uncatalyzed activation energy of reactions that life manages to perform,

$O_2$-$N_2$-$H_2O$ is probably high because it involves breaking the triple bond in $N \equiv N$ by oxidation. The reaction between $CO_2$, $N_2$, $CH_4$, and $H_2O$ (Equation (12)) also involves breaking an $N_2$ bond, so it potentially has an activation energy comparable to Reaction (13) (>300 kJ/mol).

Nitrogen fixing bacteria are the only organisms that break $N \equiv N$ bonds by chemical reduction with the aid of the nitrogenase enzyme. The literature suggests that the uncatalyzed activation energy of nitrogen fixation by reduction is ~200 kJ/mol (Hageman & Burris 1980), which is <63% of the uncatalyzed activation energy of Reaction (13). These differing energy barriers might explain why biology has managed to catalyze nitrogen fixation by reduction of $N_2$ but not by direct oxidation of $N_2$.

In summary, we speculate that life has not evolved to consume the $CO_2$-$N_2$-$CH_4$-$H_2O$ and $O_2$-$N_2$-$H_2O$ disequilibrium because these reactions are kinetically insurmountable for biology. We hypothesize that these reactions will be biochemically prohibited elsewhere on Earth-like exoplanets, which is a testable hypothesis (section 4.4).

### 4.3. Chemical disequilibrium as a biosignature or anti-biosignature

Throughout Earth's history, the available Gibbs energy of the atmosphere-ocean system varied substantially (Figure 3), and there is no one-to-one relationship between the magnitude of Gibbs energy and the presence of life. In both prebiotic and modern times, the atmosphere-ocean disequilibrium was relatively large (~1000s J/mol), so high disequilibrium alone is not a reliable sign of life. Lower disequilibrium (~100s) is also an ambiguous biosignature on its own because there were large spans of Earth's inhabited history when disequilibrium was comparable to the available Gibbs energy of Mars' atmosphere (136 J/mol) (Krissansen-Totton, et al. 2016).

However, disequilibrium is useful to determine the presence or absence of life if you know which particular species are responsible for the disequilibrium. The species causing the prebiotic and modern disequilibrium are different even though the magnitude of disequilibrium is similar. Before life appeared, atmospheric disequilibrium was dominated by $H_2$-$CO_2$, and $CO$-$H_2O$, while today the most important disequilibrium is $O_2$-$N_2$-$H_2O$.

Thus, biosignatures and anti-biosignatures arise from looking at both the magnitude of disequilibrium and how "edible" the disequilibrium gas mixture is, where "edibility" is associated with a low activation energy. An atmosphere-ocean with "edible" disequilibrium is an anti-biosignature because it is a sign that life is not consuming disequilibria that has kinetic barriers that are easily biologically surmountable (Table 2). One example is the prebiotic Earth, which likely had large amounts of "edible" free energy from the $H_2$-$CO_2$ and CO-$H_2O$ disequilibria. If chemotrophs were present, these "edible" disequilibria would mostly be destroyed.

A separate example of an anti-biosignature is Mars' atmosphere, which has a fairly large available Gibbs energy (~136 J/mol) mostly because of photochemically produced CO and $O_2$ (Krissansen-Totton, et al. 2016). This free energy could be consumed by aerobic carboxydotrophic organisms (Sholes, et al. 2019). If a substantial biosphere were present, then it would consume this "edible" free lunch because a known enzyme (aerobic CO dehydrogenase) makes CO readily consumable with an activation energy ranging ~20-95 kJ/mol (King 2013; Xie et al. 2009). Strictly speaking, then, an anti-biosignature provides an upper limit on biomass (Sholes, et al. 2019).

An atmosphere-ocean with primarily "inedible" disequilibrium (with an insurmountable activation energy barrier) is a biosignature (top right of Table 2). In this case, chemotrophs have consumed most of the "edible" free energy produced by geology or photosynthesis (if present) and have left "inedible" redox couples untouched. Some small amount of "edible" disequilibrium will always remain, because gas fluxes from the atmosphere into habitable bodies of water will be limited by the water boundary layer (Liss & Slater 1974). The magnitude of the "inedible" disequilibrium should be larger if phototrophs are present. While life has been present on Earth, the coexistence of "inedible" $CO_2$-$N_2$-$CH_4$-$H_2O$ or $O_2$-$N_2$-$H_2O$ has persisted in Earth's atmosphere-ocean system (Figure 3), and "edible" disequilibrium has been absent because of chemotrophs.

A planet very near thermodynamic equilibrium most likely does not have life (lower row of Table 2). Although chemotrophs destroy disequilibrium, they did not drive Earth's atmosphere-ocean system to complete equilibrium in the Archean. Chemotrophs on Earth produce waste gas such as $CH_4$ (Equation (4)) that ultimately contribute to disequilibria and therefore life is unable to destroy all atmospheric disequilibrium.

The difference between the upper left and lower row of Table 2 is a question of degree. The upper left represents an anti-biosignature applicable to a large disequilibrium, such as prebiotic Earth ~$10^3$ J/mol or of modern Mars ~$10^2$ J/mol. In contrast, the lower row of Table 2 is applicable to a planet such as Venus, where the near-surface temperature drives the atmosphere very close to equilibrium with a disequilibrium of 0.06 J/mol (Krissansen-Totton et al., 2016). Also in this category are giant planets, such as Jupiter, where deep convective mixing produces a gas mixture very near chemical equilibrium (~0.001 and the small disequilibrium is purely photochemical.

Some biospheres that are nutrient-limited (e.g., limited by fixed N or P) may not follow Table 2. For example, a nutrient-limited chemotrophic biosphere may not be able to consume all of the “edible” disequilibrium in the atmosphere. In this case, sizable “edible” disequilibrium might coexist with life, which contradicts the upper-left panel of Table 2. Most literature has argued that the pre-photosynthetic Earth was probably energy-limited (not nutrient-limited) (Canfield, et al. 2006; Kharecha, et al. 2005; Ward, et al. 2019), therefore it might be reasonable to expect other purely chemotrophic biospheres to be energy-limited.

**Table 2**

Chemical disequilibrium as a biosignature and anti-biosignature.

| | Primarily "edible" disequilibria (low activation energy) | Primarily "inedible" disequilibria (high activation energy) |
|---|---|---|
| Atmosphere-ocean in disequilibrium | **Anti-biosignature**<br>The presence of uneaten "edible" food should be consumed by biology. | **Biosignature**<br>Life has consumed most of the "edible" food produced by geology and photosynthetic life (if present) and has left the "inedible" food untouched. The magnitude of the "inedible" disequilibrium should be <u>larger</u> if phototrophs are present, and <u>smaller</u> if only chemotrophs are present. |
| Atmosphere-ocean near equilibrium | **Anti-biosignature**<br>Although chemotrophic life destroys disequilibrium, it is unlikely to drive a system to complete thermodynamic equilibrium. Chemotrophic metabolisms produce waste gases that are "inedible," so they leave some fraction of a planet’s disequilibrium unconsumed. Therefore, a planet near equilibrium instead will be characterized by small abiotic disequilibrium resulting from photochemistry or small volcanic fluxes, if volcanism is present. The planet is very likely uninhabited although an extremely meager, undetectable biosphere cannot be excluded. | |

There are some cases where even a productive biosphere can coexist with edible atmospheric disequilibrium. This is because there are limits to how quickly gases can be transported from the atmosphere, into the ocean where they can be consumed by life (Kharecha, et al. 2005). For example, consider a planet with a very large volcanic CO flux (e.g. 100x modern). CO could build up in this planet’s atmosphere even if CO consumers were present in an ocean, because CO transport from the atmosphere to the ocean would not be sufficient to maintain low atmospheric CO (Schwieterman, et al. 2019). While coexistence of productive biospheres and edible disequilibrium is conceivable, it might be unlikely on exoplanets, given that it probably did not occur during all of Earth's history (Figure 1).

These aforementioned caveats to Table 2 highlights the importance of inferring fluxes of gases to further evaluate disequilibrium biosignatures (Krissansen-Totton, et al. 2018c; Simoncini et al. 2013). The indicator of biology is a surface flux of gases not explained by geology, although the atmospheric composition resulting from a biological flux depends on many factors like the host star's spectrum, or volcanic outgassing rates (Segura et al. 2005). Therefore, it makes sense to infer surface fluxes of disequilibrium gases and then compare inferred fluxes to dead processes. Fluxes unexplained by dead processes are evidence for life. Detailed consideration of fluxes is beyond the scope of this paper.

### 4.4. Detecting the prebiotic Earth disequilibrium anti-biosignature

The prebiotic disequilibrium anti-biosignature is, in principle, remotely detectable on exoplanets. Strong spectral signatures of atmospheric $CO_2$, CO and $H_2O$ exist, and could be detected with reflectance or transmission spectroscopy (see Table 3 in Catling et al. (2018)). The presence of prebiotic $H_2$ could be inferred with its spectral feature at $2.12\,\mu m$, or its continuous features in the near-infrared and $< 0.08\,\mu m$. $H_2$ could also be detected by combining several spectral methods. Ultraviolet transmission spectroscopy can be used to observe hydrogen escape because hydrogen absorbs stellar Lyman-alpha. This has been done for warm Neptunes (Ehrenreich et al. 2015), and could be done for Earth sized planets with future telescopes (Fujii et al. 2018). If $CH_4$ and stratospheric $H_2O$ were ruled out with transmission spectroscopy, then the hydrogen escape must result from $H_2$ in the atmosphere.

## 5. Conclusions

Given our current knowledge of photochemistry and Earth's Hadean atmosphere, we calculate that Earth's prebiotic atmosphere was in thermodynamic chemical disequilibrium due primarily to volcanic outgassing, and that the advent of chemosynthetic life destroyed much of this disequilibrium through its metabolism. Subsequently, disequilibrium rose for the rest of Earth's history primarily because oxygenic photosynthesis maintained high $O_2$ and $N_2$ levels, directly and indirectly, respectively.

In the prebiotic era, volcanically produced $H_2$-$CO_2$ and CO-$H_2O$ were the largest contributors to the atmosphere-ocean available Gibbs energy. After the origin of life, chemotrophs consumed most of the prebiotic free energy, although the atmosphere-ocean system remained in disequilibrium because of biological waste gases: $CO_2$, $CH_4$, $N_2$ and liquid water. After the Great Oxidation Event, the magnitude of the available Gibbs energy rose, and was instead dominated by $O_2$, $N_2$ and liquid water.

Earth's history reveals a different relationship between life and atmospheric chemical disequilibrium than was first proposed by Lovelock (1965). Lovelock (1965) argued that planets with life should be in disequilibrium and that dead worlds should be near equilibrium, although we have shown that this was not true and was subtler for the first billion years of Earth history.

We suggest that chemotrophs never evolved to consume the $CO_2$-$N_2$-$CH_4$-$H_2O$ disequilibrium prior to atmospheric oxygenation and $O_2$-$N_2$-$H_2O$ disequilibrium after oxygenation because the reaction of these groups of species has insurmountable activation energy barriers. In contrast, the reactions between $H_2$ and $CO_2$ or CO and $H_2O$ have activation energy barriers that can be lowered by enzymes, so that these redox couples readily support microbial metabolisms.

The large prebiotic "edible" disequilibrium between $H_2$ and $CO_2$ or CO and $H_2O$ is therefore an anti-biosignature because these easily metabolized species should be consumed by chemotrophs. A planet that is dominated by "inedible" disequilibria such as $CO_2$-$N_2$-$CH_4$-$H_2O$ or $O_2$-$N_2$-$H_2O$ has signs of biology because these disequilibria show that life has consumed most the "edible" food produced by abiotic processes and has created "inedible" disequilibria with continuous fluxes of waste gases.

The mere detection of "edible" or "inedible" disequilibria is not a definitive sign of the presence or absence of life.  A full evaluation of disequilibria would compare inferred surface fluxes of disequilibrium gases to plausible abiotic surface fluxes, which is further work beyond the focus of the present paper.

This work was supported by the NASA Astrobiology Institute's Virtual Planetary Laboratory grant NNA13AA93A. We thank Josh Krissansen-Totton and an anonymous reviewer for helpful comments that improved our paper.

APPENDIX A. VOLCANIC OUTGASSING FLUXES

One input for the model of photochemistry coupled to a microbial ecosystem is the flux of volcanic outgassing. Here we describe how plausible prebiotic volcanic fluxes are calculated.

We assume that gases emitted by a volcanic melt achieve thermodynamic equilibrium. The reactions governing equilibrium of volcanic gases are

$$H_2O \leftrightarrow H_2 + \frac{1}{2}O_2 \tag{16}$$

$$CO_2 \leftrightarrow CO + \frac{1}{2}O_2 \tag{17}$$

$$CO_2 + 2H_2O \leftrightarrow CH_4 + 2O_2 \tag{18}$$

$$SO_2 + H_2O \leftrightarrow H_2S + \frac{3}{2}O_2 \tag{19}$$

At equilibrium, the ratios of the fugacities of volatile species (denoted $f_x$) are related to the equilibrium constant corresponding to each chemical reaction. The fugacities of each species are well approximated by magma chamber partial pressures (denoted $P_x$) because we consider low pressures and high temperatures (5 bar and 1473 K, following Holland (1984)), so non-ideal corrections can be ignored.

$$K_1 = \frac{f_{H_2} f_{O_2}^{0.5}}{f_{H_2O}} \approx \frac{P_{H2} f_{O_2}^{0.5}}{P_{H_2O}} \tag{20}$$

$$K_2 = \frac{f_{CO} f_{O_2}^{0.5}}{f_{CO_2}} \approx \frac{P_{CO} f_{O_2}^{0.5}}{P_{CO_2}} \tag{21}$$

$$K_3 = \frac{f_{CH_4} f_{O_2}^{2}}{f_{CO_2} f_{H_2O}^{2}} \approx \frac{P_{CH_4} f_{O_2}^{2}}{P_{CO_2} P_{H_2O}^{2}} \tag{22}$$

$$K_4 = \frac{f_{H_2S} f_{O_2}^{1.5}}{f_{SO_2} f_{H_2O}} \approx \frac{P_{H_2S} f_{O_2}^{1.5}}{P_{SO_2} P_{H_2O}} \tag{23}$$

We calculate equilibrium constants for temperature $T = 1473$ K using the NASA thermodynamic database (Burcat & Ruscic 2005). Additionally, we estimate the oxygen fugacity ($f_{O_2}$) of prebiotic volcanic gases by a linear regression through data obtained from Aulbach and Stagno (2016) (Figure 5). We take $\log(f_{O_2}) = \mathrm{FMQ} - 1.48$ at 4.0 Ga as a prebiotic value. At the temperatures and pressures we consider ( $T = 1473\text{ K}$ and $P = 5\text{ bar}$), $\log(\text{FMQ}) = -8.47$, our Gibbs energy calculations are fairly insensitive to the chosen oxygen fugacity at 4 Ga. Changing the oxygen fugacity by 1 log unit changes our calculated Gibbs energy results by a factor of ~2 (See Appendix B.3).

We also assume that the ratio of carbon to hydrogen ( $\chi_{\text{C}}$), and sulfur to hydrogen ( $\chi_{\text{S}}$) in volcanic gases has remained constant through Earth's history. This is a reasonable assumption because these ratios depend most on the pressure of degassing (Gaillard & Scaillet 2014), i.e., the atmospheric pressure into which the gases are released, and atmospheric pressure has probably has not changed by orders of magnitude over Earth's history (Som et al. 2012).

$$\frac{P_{\text{CO}_2} + P_{\text{CO}} + P_{\text{CH}_4}}{P_{\text{H}_2} + P_{\text{H}_2\text{O}} + 2P_{\text{CH}_4} + P_{\text{H}_2\text{S}}} = \chi_{\text{C}} \tag{24}$$

$$\frac{P_{\text{H}_2\text{S}} + P_{\text{SO}_2}}{P_{\text{H}_2} + P_{\text{H}_2\text{O}} + 2P_{\text{CH}_4} + P_{\text{H}_2\text{S}}} = \chi_{\text{S}} \tag{25}$$

We calculate $\chi_{\text{C}}$ and $\chi_{\text{S}}$ using modern values of total volcanic outgassing which we take from Catling and Kasting (2017), Chapter 7 (their Table 7.1). The total fluxes of hydrogen, carbon and sulfur are given by summing all species weighted by the number of atoms each species contains (e.g. $F_{\text{H}_2} + F_{\text{H}_2\text{O}} + 2F_{\text{CH}_4} + F_{\text{H}_2\text{S}} = F_{\text{hydrogen}}^{\text{mod}}$). The ratios of total fluxes are then calculated in the following way:

$$\chi_{\text{C}} = \frac{F_{\text{carbon}}^{\text{mod}}}{F_{\text{hydrogen}}^{\text{mod}}} \tag{26}$$

$$\chi_{\text{S}} = \frac{F_{\text{sulfur}}^{\text{mod}}}{F_{\text{hydrogen}}^{\text{mod}}} \tag{27}$$

Modern fluxes, and ratios are given in Table 3. We also assume that the partial pressures sum to the magma chamber total pressure:

$$P_{\text{H}_2} + P_{\text{H}_2\text{O}} + P_{\text{CH}_4} + P_{\text{H}_2\text{S}} + P_{\text{SO}_2} + P_{\text{CO}_2} + P_{\text{CO}} = P \tag{28}$$

Equations (20)-(25) and (28) are a system of 7 equations with 7 unknown partial pressures ( $P_{\text{H}_2}$, $P_{\text{H}_2\text{O}}$, etc.), which can be solved with some algebraic manipulation.

With the partial pressures in hand, we can calculate plausible prebiotic volcanic outgassing fluxes with another system of equations:

$$\frac{F_{H_2}}{F_{H_2O}} = \frac{P_{H_2}}{P_{H_2O}} \tag{29}$$

$$\frac{F_{CO}}{F_{CO_2}} = \frac{P_{CO}}{P_{CO_2}} \tag{30}$$

$$\frac{F_{CH_4}}{F_{CO_2}} = \frac{P_{CH_4}}{P_{CO_2}} \tag{31}$$

$$\frac{F_{H_2S}}{F_{SO_2}} = \frac{P_{H_2S}}{P_{SO_2}} \tag{32}$$

$$F_{H_2} + F_{H_2O} + 2F_{CH_4} + F_{H_2S} = F_{\text{hydrogen}} \tag{33}$$

$$F_{CO_2} + F_{CO} + F_{CH_4} = F_{\text{carbon}} \tag{34}$$

$$F_{SO_2} + F_{H_2S} = F_{\text{sulfur}} \tag{35}$$

The first four equations come from assuming that ratios between volcanic fluxes are equal to the corresponding ratios of the partial pressures. The final three equations are sums of the total hydrogen, carbon and sulfur fluxes weighted by the number of atoms in each species.

The total flux of each species (e.g. $F_{\text{hydrogen}}$) on the prebiotic Earth is uncertain and depends on the tectonic regime and its association with outgassing. If the Earth lacked plate tectonics and was in a "stagnant lid" regime, then heat fluxes could have been the same as modern fluxes despite a much warmer mantle (Korenaga 2009). On the other hand, if plate tectonics or some similar precursor was active in the Hadean, heat fluxes on the 4 Ga Earth could have been 5 times higher than today's fluxes (Zahnle, et al. 2001). Volcanic outgassing can be related to heat flow with a power law.

$$F_x = F_x^{\text{mod}} Q^n \tag{36}$$

Here, $Q$ is heat flow normalized to present, $F_x$ is the outgassing flux of species $x$, and $n$ is between 1 and 2 (Krissansen-Totton, et al. 2018a). Taking 5 and 1 for upper and lower bounds for heat flow ($Q$) at 4 Ga, respectively, and conservatively taking $n = 2$ gives outgassing rates between 1 and 25 times modern outgassing rates. We adopt this large range here to calculate $F_{\text{hydrogen}}$, $F_{\text{carbon}}$, and $F_{\text{sulfur}}$:

$$F_{\text{hydrogen}} = CF_{\text{hydrogen}}^{\text{mod}} \tag{37}$$

$$F_{\text{carbon}} = CF_{\text{carbon}}^{\text{mod}} \tag{38}$$

$$F_{\text{sulfur}} = CF_{\text{sulfur}}^{\text{mod}} \tag{39}$$

Here, $C$ is the outgassing multiplier, which we vary between 1 and 25 to capture the most likely outgassing scenarios on the prebiotic Earth. Equations (29) - (35) are a system of 7 linear equations with 7 unknown volcanic fluxes (e.g. $F_{H_2}$), which can be reorganized and solved with matrix inversion. We solve this system for outgassing parameters ( $C$ ) between 1 and 25, which yields a range of outgassing fluxes for each of each of the 7 species.

## APPENDIX B. PHOTOCHEMICAL MODELING AND GIBBS ENERGY MINIMIZATION

### B.1. Photochemical Modeling

Table 4 and Table 5 contains most of the boundary conditions used for modeling the prebiotic and chemosynthetic atmospheres respectively with the *Atmos* photochemical model. All species that are not listed in Table 4 and Table 5 that are in the *Atmos* code, have deposition velocities set to zero.

Photochemistry depends on the temperature and $H_2O$ mixing ratio in the atmosphere (Figure 6). We acquire temperature and $H_2O$ profiles by coupling the *Atmos* photochemical code with the *Atmos* 1-D radiative-convective climate model. This is done by running the photochemical code, then using its output as input for the climate model. The temperature and $H_2O$ output of the climate model is then used as input for the photochemical code. This coupling is continued until convergence is reached. We only couple the photochemical-climate code for the lowest volcanic outgassing scenario ( $C = 1$) in the prebiotic case and use the resulting $H_2O$ and temperature profiles for all simulations. Using the climate code for each simulation independently did not change the results significantly. The temperature and $H_2O$ profile used in this study is shown in Figure 6.

Currently, the open-source version of *Atmos* has several rate constants which are inappropriately "zeroed." We updated these rate constants to their proper values following (Harman et al. 2015). Table 6 shows a list of the updated rate constants.

All of our models include the modern production rate of NO from lightning, although this does not affect our results significantly. Every simulation uses the Sun's spectrum at 4 Ga calculated using the "Youngsun" routine (Claire et al. 2012).

### B.2. Chemical disequilibrium calculation with Gibbs energy minimization

For each modeled prebiotic and biotic atmosphere, we calculate the atmosphere-ocean chemical disequilibrium with Gibbs energy minimization. Figure 7 and Figure 8 illustrate this calculation for the lowest volcanic outgassing scenario (outgassing multiplicative factor $C = 1$) for the prebiotic and biotic Earth respectively. For both figures, the blue bars are the initial or observed concentration of the atmosphere that we generated with the *Atmos* photochemical code. The red bars are the concentrations for the atmosphere-ocean system at chemical equilibrium.

The chemical reactions that contribute most to the chemical disequilibrium are apparent in Figure 7 and Figure 8. The main disequilibria in the prebiotic atmosphere are $H_2$-$CO_2$ and CO-$H_2O$. Figure 7 shows that the atmosphere at equilibrium has much less $H_2$ and CO than the initial state. Additionally, the chemotrophic Earth's main disequilibrium was $CO_2$-$N_2$-$CH_4$-$H_2O$, which can be seen in Figure 8 because the equilibrium state has much less $CH_4$ than the initial state.

### B.3. Sensitivity of chemical disequilibrium calculations to oxygen fugacity

Figure 9 shows that chemical disequilibrium is fairly insensitive to the mantle oxygen fugacity. Chemical disequilibrium, as measured by Gibbs energy, is plotted for the lowest volcanic outgassing scenario ( $C = 1$) as a function of oxygen fugacity. Changing the oxygen fugacity by 1 log unit changes the calculated Gibbs energy results by a factor of ~2 (Figure 9). This effect is small compared to the uncertainty in volcanic outgassing rates, so it seems reasonable to ignore it.

## APPENDIX C. UNCATALYZED ACTIVATION ENERGY OF NITROGEN FIXATION

We suggest that life did not evolve to consume the $O_2$-$N_2$-$H_2O$ and $CO_2$-$N_2$-$CH_4$-$H_2O$ disequilibria because reactions of gases in these disequilibria have biologically insurmountable kinetic barriers. To substantiate this argument, we compare the uncatalyzed activation energy of $O_2$-$N_2$-$H_2O$ (>316 kJ/mol) to the uncatalyzed activation energy of nitrogen fixation reduction, because nitrogen fixation by reduction is arguably the most kinetically difficult reaction that biology has managed to catalyze. The net nitrogen fixation by reduction reaction is

$$3H_2 + N_2 \rightarrow 2NH_3 \tag{40}$$

Table 7 summarizes literature data on the uncatalyzed activation energy of this reaction. Estimates range from 150 kJ/mol to 200 kJ/mol, although we plot the 200 kJ/mol value in Figure 4b.

**Appendix Figures**

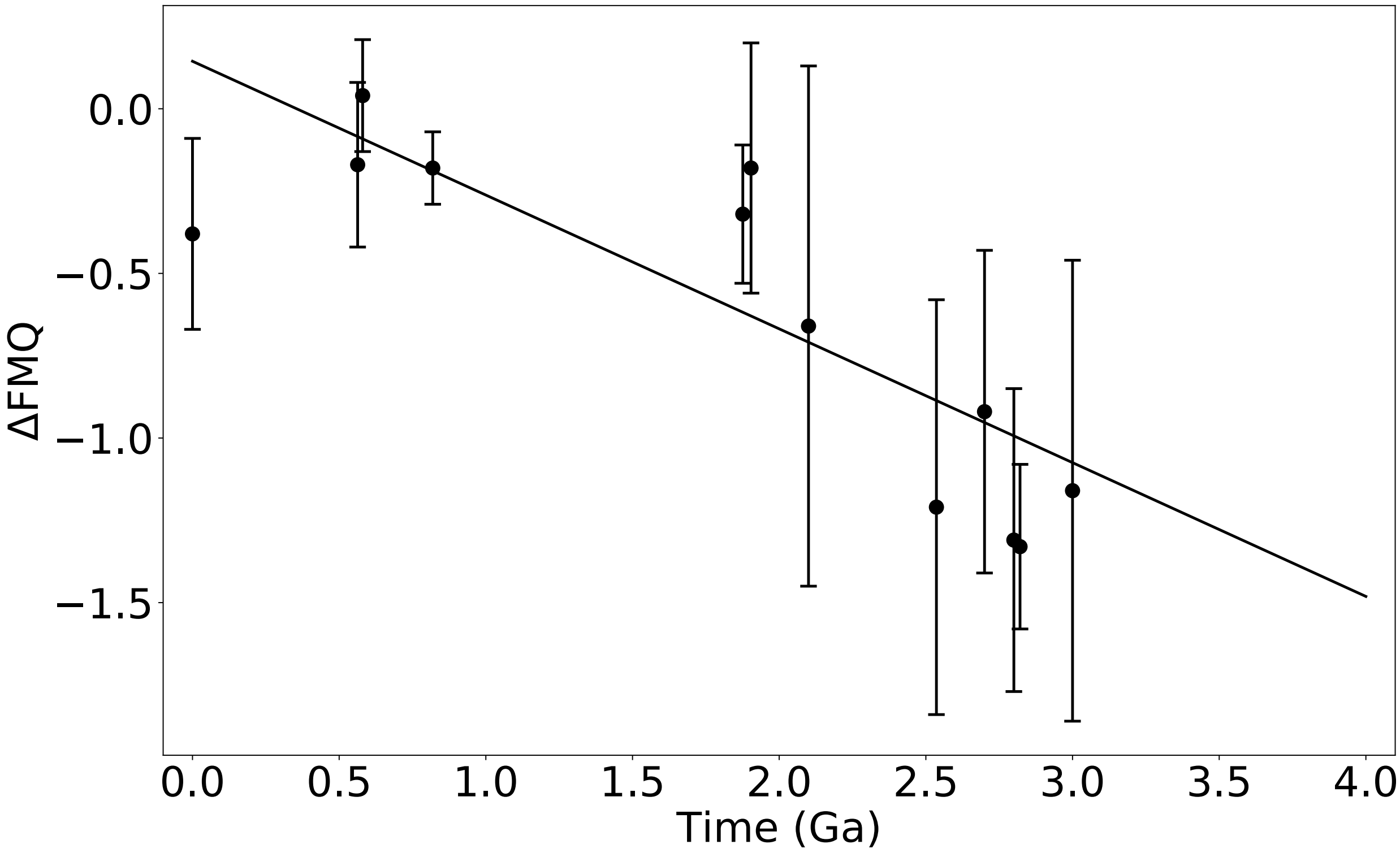

Figure 5: Weighted linear fit of mantle redox proxies from Aulbach and Stagno (2016). At 4 Ga, the linear fit predicts $\log(f_{O2}) = \mathrm{FMQ} - 1.48$.

**Table 3**

Modern mantle-sourced volcanic outgassing fluxes and ratios.

| Modern Volcanic Fluxes (Tmol/yr) | | | | | | | Total Modern Fluxes (Tmol/yr) | | | Ratios | |
|---|---|---|---|---|---|---|---|---|---|---|---|
| $CO_2$ | $H_2O$ | $SO_2$ | $H_2$ | CO | $CH_4$ | $H_2S$ | $F_{\text{hydrogen}}^{\text{mod}}$ | $F_{\text{carbon}}^{\text{mod}}$ | $F_{\text{sulfur}}^{\text{mod}}$ | $\chi_C$ | $\chi_S$ |
| 8.5 | 95 | 1.8 | 2.0 | 0.25 | 0 | 0.03 | 97.03 | 8.75 | 1.83 | 0.090 | 0.019 |

Note – Fluxes of $CO_2$, $H_2O$, $SO_2$, and $H_2S$ are from Catling and Kasting (2017) p. 203 and p. 212. Fluxes of $H_2$, CO, and $CH_4$ are calculated using equilibrium (e.g., Equation (20) with Equation (29)) and assuming $T = 1473\,\text{K}$, $P = 5\,\text{bar}$, and $\log(f_{O2}) = \text{FMQ}$. The total modern fluxes ( $F_x^{mod}$), and ratios $\chi_C$ and $\chi_S$ are calculated using the modern outgassing fluxes. Methods for this calculation are detailed in the text.

**Table 4**

Prebiotic boundary conditions.

| Chemical Species | Deposition Velocity (cm $s^{-1}$) | Mixing Ratio | Flux (molecules $cm^{-2}$ $s^{-1}$) |
|---|---|---|---|
| O | 1.00E+00 | - | - |
| $O_2$ | 1.40E-04 | - | - |
| $H_2O$ | 0 | - | - |
| H | 1.00E+00 | - | - |
| OH | 1.00E+00 | - | - |
| $HO_2$ | 1.00E+00 | - | - |
| $H_2O_2$ | 2.00E-01 | - | - |
| $H_2$ | 0 | - | variable |
| CO | 1.00E-08 | - | variable |
| HCO | 1.00E+00 | - | - |
| $H_2CO$ | 2.00E-01 | - | - |
| $CH_4$ | 0 | - | 0.00E+00 |
| $CH_3$ | 1.00E+00 | - | - |
| $C_2H_6$ | 0 | - | - |
| NO | 3.00E-04 | - | - |
| $NO_2$ | 3.00E-03 | - | - |
| HNO | 1.00E+00 | - | - |
| $O_3$ | 7.00E-02 | - | - |
| $HNO_3$ | 2.00E-01 | - | - |
| $H_2S$ | 2.00E-02 | - | variable |
| $SO_3$ | 0 | - | - |
| $S_2$ | 0 | - | - |
| HSO | 1.00E+00 | - | - |
| $H_2SO_4$ | 1.00E+00 | - | - |
| $SO_2$ | 1.00E+00 | - | variable |
| SO | 0 | - | - |
| $SO_4$ aerosol | 1.00E-02 | - | - |
| $S_8$ aerosol | 1.00E-02 | - | - |
| hydrocarbon aerosol | 1.00E-02 | - | - |
| $CO_2$ | - | 2.00E-01 | - |
| $N_2$ | - | 7.50E-01 | - |

Note - Species included in the photochemical scheme with a deposition velocity and flux of 0 include: N, $C_3H_2$, $C_3H_3$, $CH_3C_2H$, $CH_2CCH_2$, $C_3H_5$, $C_2H_5CHO$, $C_3H_6$, $C_3H_7$, $C_3H_8$, $C_2H_4OH$, $C_2H_2OH$, $C_2H_5$, $C_2H_4$, CH, $CH_3O_2$, $CH_3O$, $CH_2CO$, $CH_3CO$, $CH_3CHO$, $C_2H_2$, $(CH_2)_3$, $C_2H$, $C_2$, $C_2H_3$, HCS, $CS_2$, CS, OCS, S, and HS. Here, deposition velocities follow those used by Kharecha, et al. (2005) and Schwieterman, et al. (2019).

**Table 5**

Boundary conditions for the chemotrophic ecosystem model.

| Chemical Species | Deposition Velocity (cm $s^{-1}$) | Mixing Ratio | Flux (molecules $cm^{-2}$ $s^{-1}$) |
|---|---|---|---|
| O | 1.00E+00 | - | - |
| $O_2$ | 1.40E-04 | - | - |
| $H_2O$ | 0 | - | - |
| H | 1.00E+00 | - | - |
| OH | 1.00E+00 | - | - |
| $HO_2$ | 1.00E+00 | - | - |
| $H_2O_2$ | 2.00E-01 | - | - |
| $H_2$ | - | variable | - |
| CO | 1.20E-04 | - | variable |
| HCO | 1.00E+00 | - | - |
| $H_2CO$ | 2.00E-01 | - | - |
| $CH_4$ | - | variable | - |
| $CH_3$ | 1.00E+00 | - | - |
| $C_2H_6$ | 0 | - | - |
| NO | 3.00E-04 | - | - |
| $NO_2$ | 3.00E-03 | - | - |
| HNO | 1.00E+00 | - | - |
| $O_3$ | 7.00E-02 | - | - |
| $HNO_3$ | 2.00E-01 | - | - |
| $H_2S$ | 2.00E-02 | - | variable |
| $SO_3$ | 0 | - | - |
| $S_2$ | 0 | - | - |
| HSO | 1.00E+00 | - | - |
| $H_2SO_4$ | 1.00E+00 | - | - |
| $SO_2$ | 1.00E+00 | - | variable |
| SO | 0 | - | - |
| $SO_4$ aerosol | 1.00E-02 | - | - |
| $S_8$ aerosol | 1.00E-02 | - | - |
| hydrocarbon aerosol | 1.00E-02 | - | - |
| $CO_2$ | - | 2.00E-01 | - |
| $N_2$ | - | 7.50E-01 | - |

Note - Species included in the photochemical scheme with a deposition velocity and flux of 0 include: N, $C_3H_2$, $C_3H_3$, $CH_3C_2H$, $CH_2CCH_2$, $C_3H_5$, $C_2H_5CHO$, $C_3H_6$, $C_3H_7$, $C_3H_8$, $C_2H_4OH$, $C_2H_2OH$, $C_2H_5$, $C_2H_4$, CH, $CH_3O_2$, $CH_3O$, $CH_2CO$, $CH_3CO$, $CH_3CHO$, $C_2H_2$, $(CH_2)_3$, $C_2H$, $C_2$, $C_2H_3$, HCS, $CS_2$, CS, OCS, S, and HS. Here, deposition velocities follow those used by Kharecha, et al. (2005) and Schwieterman, et al. (2019).

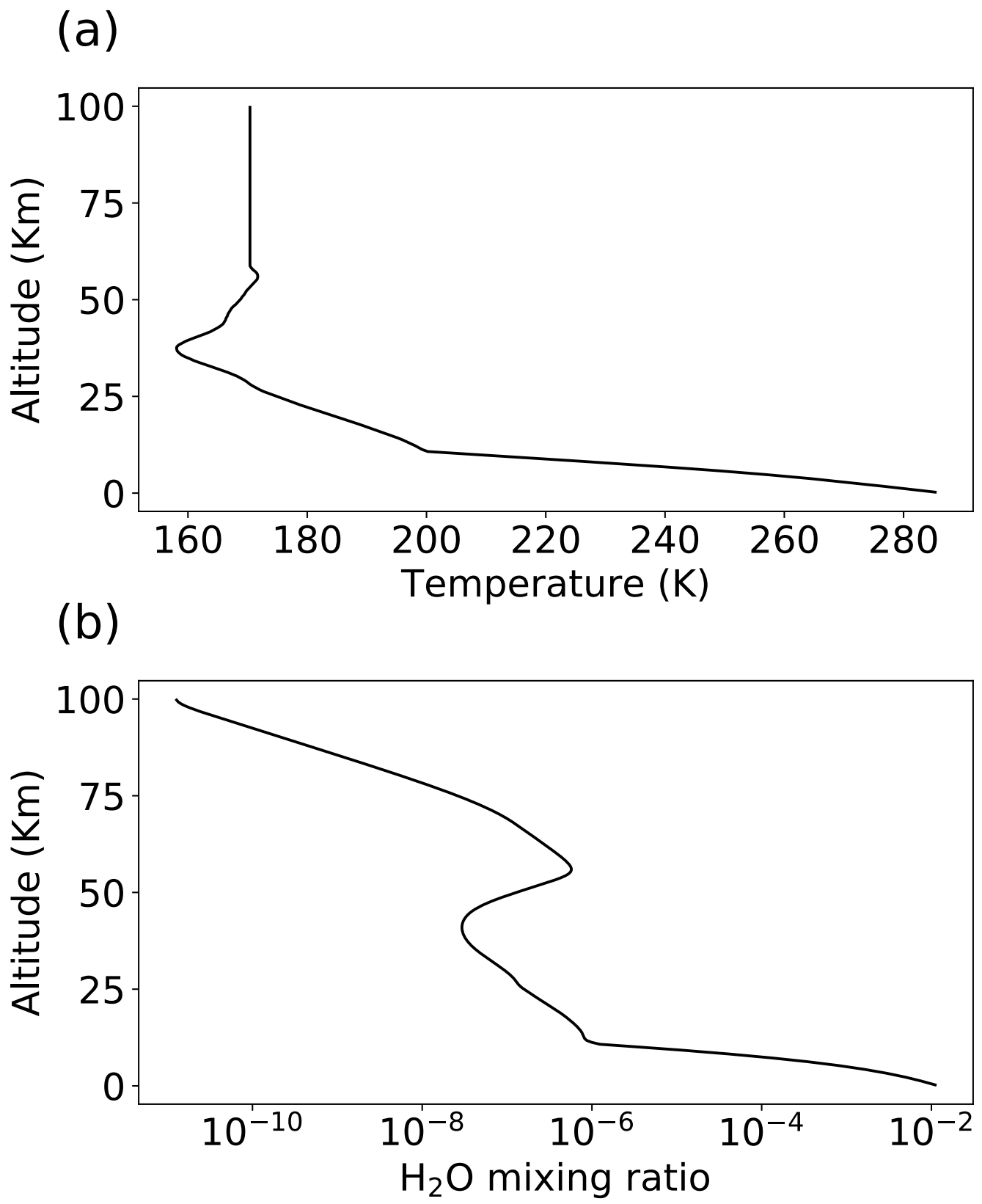

Figure 6: (a) Temperature profile and (b) $H_2O$ profile used for every simulation in this study.

**Table 6**

Updated reaction rates for *Atmos*.

| Rx # | Reaction | Rate (cm$^3$ s$^{-1}$) |
|---|---|---|
| 61 | $^3CH_2 + H_2 \rightarrow CH_3 + H$ | $5\times10^{-14}$ |
| 62 | $^3CH_2 + CH_4 \rightarrow CH_3 + CH_3$ | $6.1\times10^{12}\cdot e^{-5051/T}$ |
| 116 | $SO + HO_2 \rightarrow SO_2 + OH$ | $2.8\times10^{-11}$ |
| 123 | $HSO_3 + OH \rightarrow H_2O + SO_3$ | $1.0\times10^{-11}$ |
| 124 | $HSO_3 + H \rightarrow H_2 + SO_3$ | $1.0\times10^{-11}$ |
| 125 | $HSO_3 + O \rightarrow OH + SO_3$ | $1.0\times10^{-11}$ |
| 130 | $HS + O_2 \rightarrow OH + SO$ | $4.0\times10^{-19}$ |
| 143 | $HS + H_2CO \rightarrow H_2S + HCO$ | $1.7\times10^{-11}\cdot e^{-800/T}$ |
| 163 | $SO_2 + HO_2 \rightarrow SO_3 + OH$ | $1.0\times10^{-18}$ |
| 169 | $S + CO_2 \rightarrow SO + CO$ | $1.0\times10^{-20}$ |
| 170 | $SO + HO_2 \rightarrow HSO + O_2$ | $2.8\times10^{-11}$ |
| 174 | $HSO + NO \rightarrow HNO + SO$ | $1.0\times10^{-15}$ |

Note – All updated reaction rates are taken from Harman, et al. (2015). Harman, et al. (2015) incorrectly lists the rate for Rx #169. Here, $T$ is temperature in Kelvin.

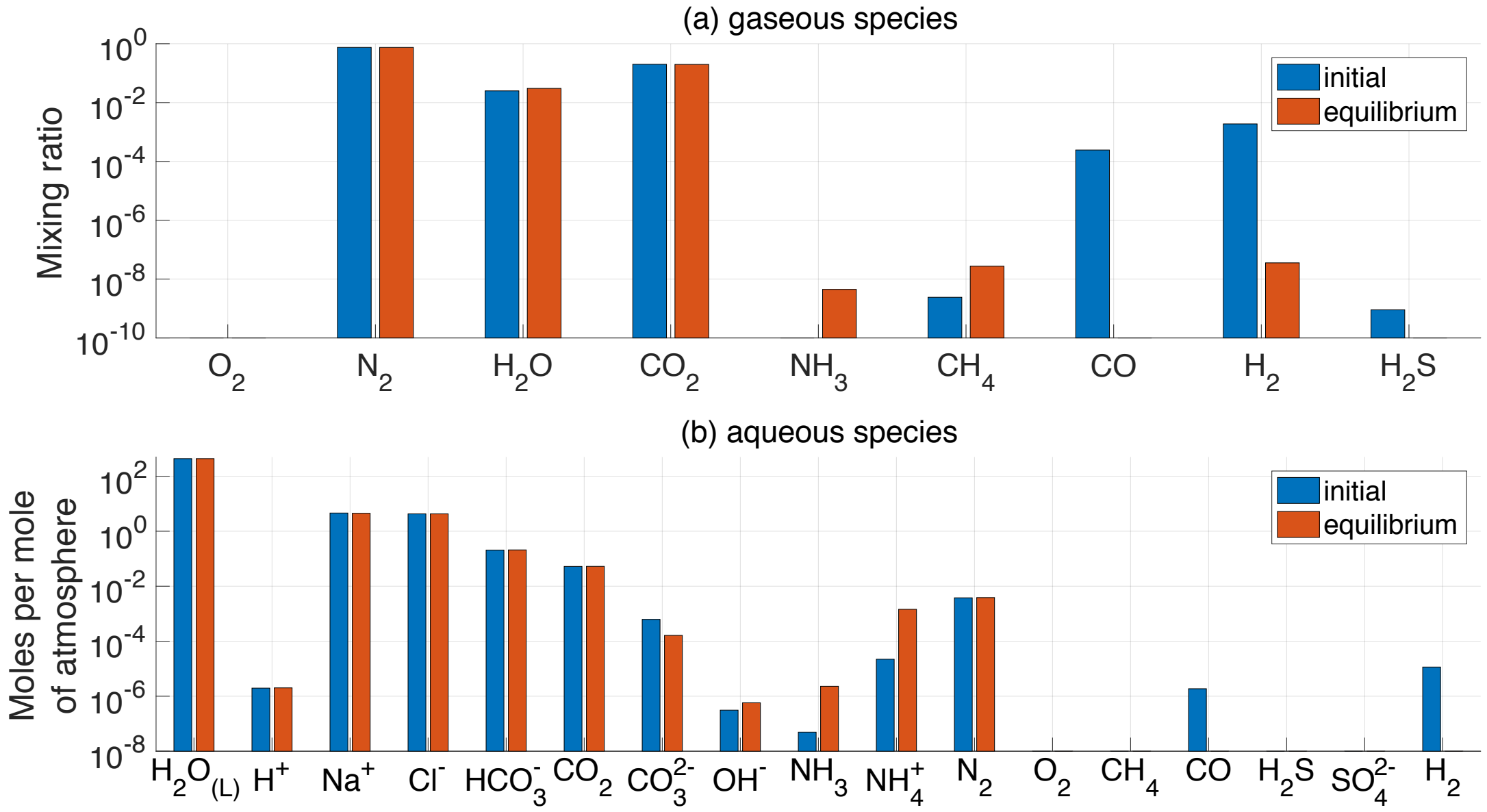

Figure 7: Atmosphere-ocean disequilibrium calculation for the prebiotic Earth (minimum outgassing scenario). Blue bars show the modeled atmosphere and ocean composition. Red bars show what happens to the species when thermodynamic equilibrium is imposed. (a) Shows all gas phase species, whereas (b) shows all aqueous species.

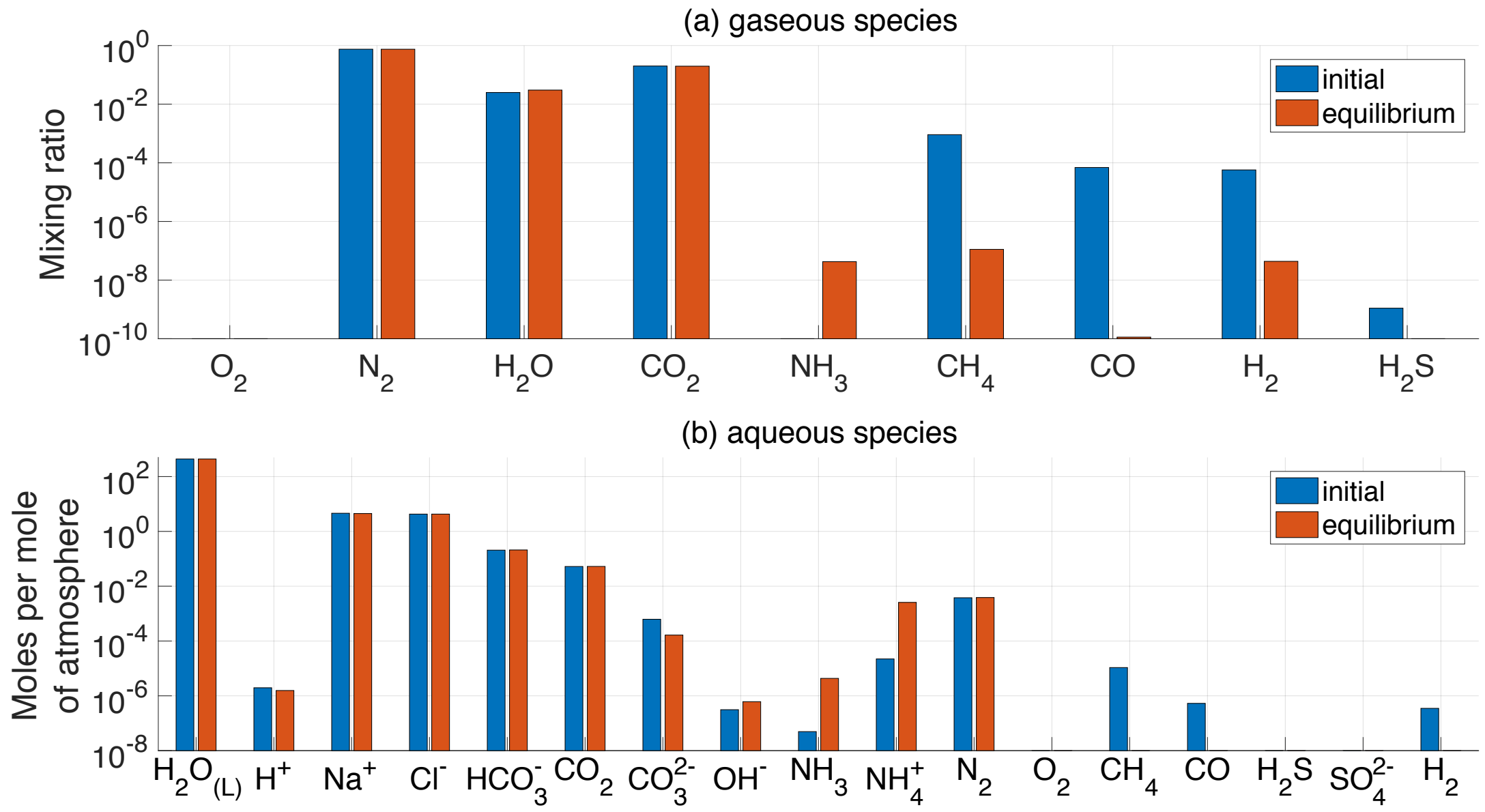

Figure 8: Atmosphere-ocean disequilibrium calculation for the chemotrophic Earth (minimum outgassing scenario). Blue bars show the modeled atmosphere and ocean composition. Red bars show what happens to the species when thermodynamic equilibrium is imposed. (a) Shows all gas phase species, whereas (b) shows all aqueous species.

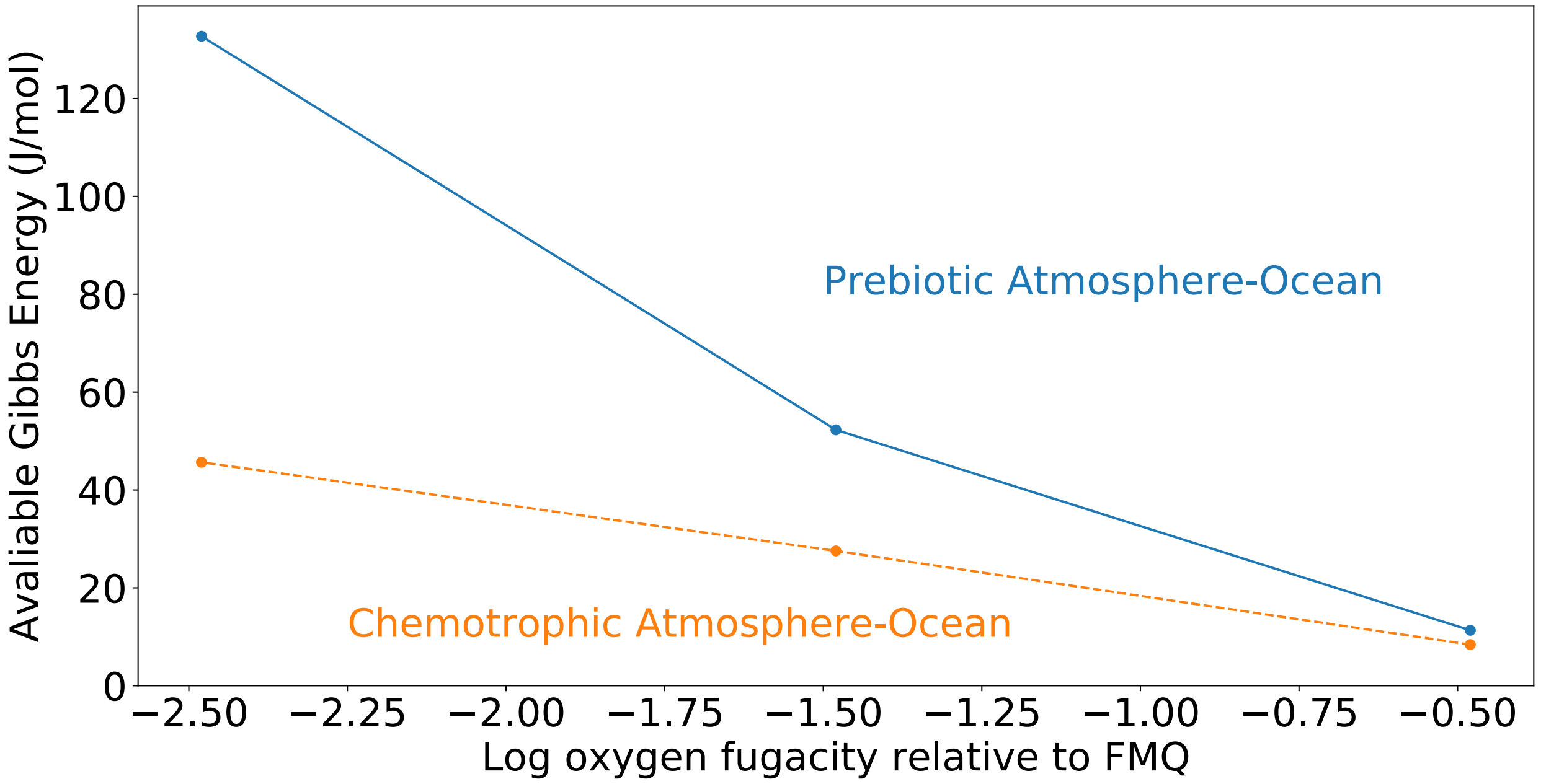

Figure 9: The effect of oxygen fugacity on the calculated available Gibbs energy for a volcanic outgassing coefficient $C = 1$ for prebiotic Earth and after the advent of a chemotrophic

biosphere. A change in 1 log unit in oxygen fugacity changes the calculated avaliable Gibbs energy by a factor of ~2.

**Table 7**

Literature values for the activation energy of nitrogen fixation by chemical reduction.

| Catalytic process | Activation Energy | Reference | Comments |
|---|---|---|---|
| With no catalyst | 200 kJ/mol | (Gutschick 1982), p. 137 | This is the Gibbs energy difference between $H_2$ and $N_2$ and the molecule $N_2H_2$ in the gas phase. $N_2H_2$ is not a step in nitrogen fixation, so this may be artificial. |
| | 150 kJ/mol | (Hageman & Burris 1980), p. 281-282 | This is the Gibbs energy difference between $H_2$ and $N_2$ and the molecule $N_2H_2$ in the aqueous phase. |
| | 150 kJ/mol | (Ljones 1979) | They claim that the activation can be understood by the reaction of $H_2$ and $N_2$ to $N_2H_2$. |
| With enzyme | 30 kJ/mol | (Andersen & Shanmugam 1977) | Between temperatures 20 and 35 °C in vivo. |
| | 60 kJ/mol | (Hardy et al. 1968) | Between temperatures 20 and 35 °C in vitro. |
| | 61 kJ/mol | (Burns 1969) | Above 21 °C. |
| | 163 kJ/mol | (Burns 1969) | Below 21 °C. |
| With non-biological catalyst | 103 kJ/mol | (Appl 1999) | On an iron surface. |
| | 27 - 60 kJ/mol | (Dahl et al. 2000) | Activation energy of $N_2$ dissociation on Ru catalyst. |
| | 131 kJ/mol | (Dahl, et al. 2000) | Calculations of $N_2$ dissociation on Ru catalyst. |
| | 101 kJ/mol | (Dahl, et al. 2000) | Supersonic molecular beam experiments. |
| | 100 - 200 kJ/mol | (Dahl, et al. 2000) | Ammonia synthesis over stepped Ru catalyst. |